\documentclass[11pt,draftclsnofoot,onecolumn]{IEEEtran} 

\usepackage{epsfig}
\usepackage{epstopdf}
\usepackage{amsmath}
\usepackage{amsfonts}
\usepackage{amssymb}
\usepackage{ amsthm}
\usepackage{float}  
\usepackage{subfig}
\usepackage{arydshln}

\renewcommand{\IEEEQED}{\IEEEQEDopen}
\newenvironment{definition}[1][Definition]{\begin{trivlist}
\item[\hskip \labelsep {\bfseries #1}]}{\end{trivlist}}

\newcommand{\beqn}{\begin{eqnarray}}
\newcommand{\eeqn}{\end{eqnarray}}
\newcommand{\beqnn}{\begin{eqnarray*}}
\newcommand{\eeqnn}{\end{eqnarray*}}
\newcommand{\vv}[1]{#1} 
\newcommand{\defeq}{\triangleq} 
\newcommand{\obliquefrac}[2]{\ensuremath{ {\raise0.7ex\hbox{$#1$} \!\mathord{\left/{\vphantom {#1 #2}}\right.\kern-\nulldelimiterspace}\!\lower0.7ex\hbox{$#2$}}}}
\newcommand{\Cbar}{\ensuremath{\overline{\textsc{C}} }}
\newcommand{\mbb}[1]{\ensuremath{\mathbb #1}}
\newcommand{\scsc}[1]{\ensuremath{ \scriptscriptstyle #1}}
\newcommand{\st}[1]{\ensuremath{ \textsc #1}}

\newcommand{\ceilp}[1]{\ensuremath{\big\lceil #1 \big\rceil}}

\newcommand{\myArrow}{\ensuremath{ \mathop{\blacktriangleleft}{\!\!\!\textrm{\textbf{=\!=\!=}}}   }}
\newcommand{\f}[1]{\fbox{#1}}

\renewcommand{\phi}{\ensuremath{\varphi}}
\def\H{{\mathrm{H}}}

\def\sH{{\mathcal{A}}}

\def\sV{{\mathcal{R}}}

\def\bH{{\mbb{H}}}
\def\V{{\mathrm{V}}}
\def\Range{{\mathfrak{R}}}
\def\cconv{{\mathcal{C}}}
\def\po{{p_{\circ}}}
\def\At{{A}} 
\def\0{{\underline{0}}}
\def\dcz{{\ensuremath{\oslash}}}
\def\1{{\underline{1}}}
\def\q{{\mathfrak{q}}}

\floatstyle{ruled} 
\newfloat{algorithm}{h}{alg}
\floatname{algorithm}{Algorithm}
\floatstyle{ruled} 
\newfloat{pcode}{h}{}
\floatname{pcode}{Psuedo-code}

\begin{document}
\title{{~\\[-20pt] \Huge{\textsc{Generic Feasibility of Perfect Reconstruction with Short FIR Filters in Multi-channel Systems}}}}
\author{Behzad~Sharif,\textsuperscript{*}~\IEEEmembership{Member,~IEEE,} and~Yoram~Bresler,~\IEEEmembership{Fellow,~IEEE}\thanks{
B. Sharif was with the Department of Electrical and Computer Engineering and Coordinated Science Laboratory, University of Illinois at Urbana-Champaign, Urbana, IL 61801, USA. He is now with the Biomedical Imaging Research Institute, Cedars-Sinai Medical Center, Los Angeles, CA 90048, USA (e-mail: \mbox{behzad.sharif@cshs.org}). 
Y. Bresler is with the Department of Electrical and Computer Engineering and Coordinated Science Laboratory, University of Illinois at Urbana-Champaign, Urbana, IL 61801, USA (e-mail: \mbox{ybresler@illinois.edu}). 
This work was supported by National Science Foundation under award CCF-1018789, and graduate fellowships from the Computation Science and Engineering program and the Beckman Institute, University of Illinois.}}
\markboth{Submitted to IEEE Trans.~Signal Processing, March~2011}{}

\maketitle
\vspace{-0.35in}
\begin{abstract}
We study the feasibility of short finite impulse response (FIR) synthesis for perfect reconstruction (PR) in generic FIR filter banks. 
Among all PR synthesis banks, we focus on the one with the minimum filter length. 
For filter banks with oversampling factors of at least two, we provide prescriptions for the shortest filter length of the synthesis bank that would guarantee PR almost surely. The prescribed length is as short or shorter than the analysis filters and has  an approximate inverse relationship with the oversampling factor. Our results are in form of necessary and sufficient statements that hold generically, hence only fail for elaborately-designed nongeneric examples. 
We provide extensive numerical verification of the theoretical results and demonstrate that the gap between the derived filter length prescriptions and the true minimum is small.  
The results have potential applications in synthesis FB design problems, where the analysis bank is given, and for analysis of fundamental limitations in blind signals reconstruction from data collected by unknown subsampled multi-channel systems.
\end{abstract}


\begin{center} 
\bfseries EDICS Category: DSP-BANK, DSP-RECO, DSP-SAMP, SAM-MCHA
\end{center}


\section{Introduction}\label{sec::intro}

\begin{figure}[b!]
\vspace{-0.15in}
\centering
\subfloat[]{\hspace{-17pt}\includegraphics[trim=0mm 2mm 0mm 0mm, clip, width=0.52 \textwidth]{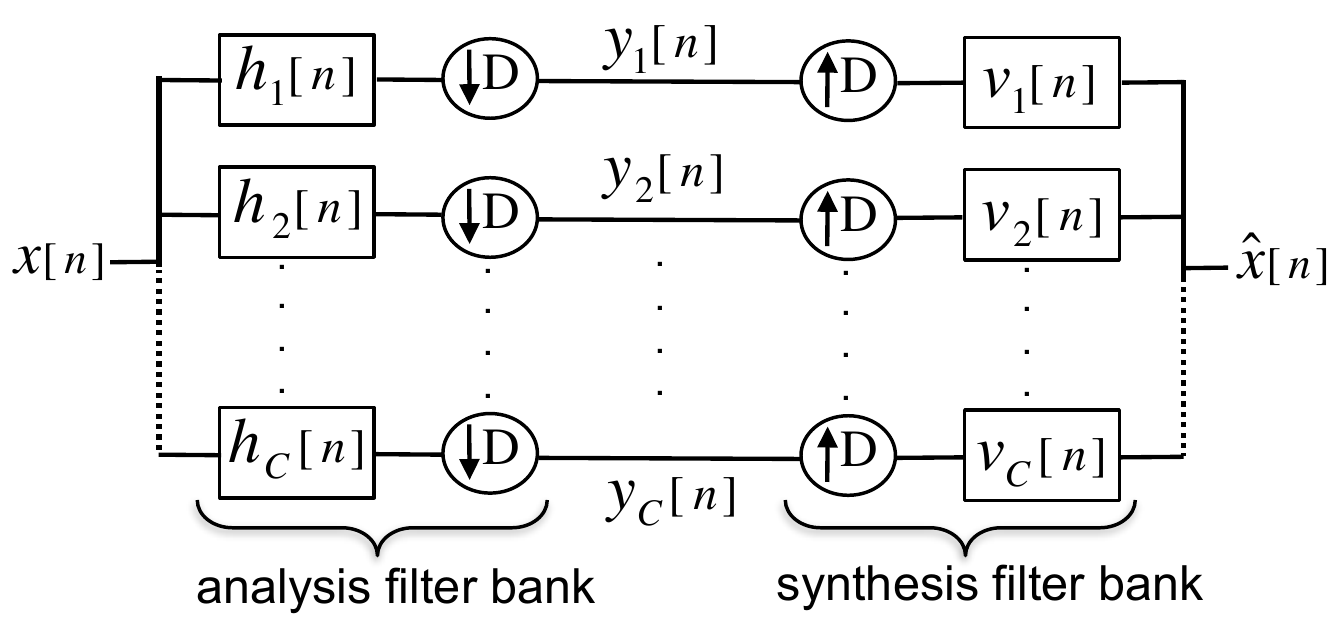}}
\subfloat[]{\includegraphics[trim=-1mm -12mm 3mm 0mm, clip, width=0.56 \textwidth]{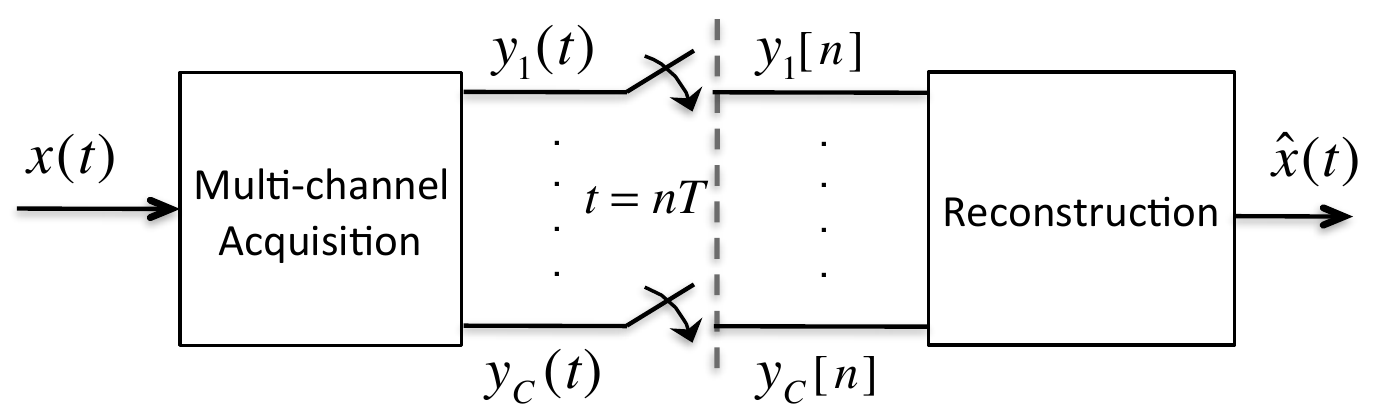}} 
\vspace{-0.15in}
\caption{\small (a) $C$-channel Filter Bank with $D$-fold subsampling.  The ratio $C/D$ is referred to as the oversampling factor. 
(b) Model for multi-channel sampling and reconstruction. 
}
\label{fig::multichannel}
\vspace{-0.2in}
\end{figure}

\vspace{-0.1in}
\subsection{Oversampled filter banks}

Filter banks with perfect reconstruction (PR), or near PR, are the most ubiquitous signal processing structure in multi-rate digital systems with applications in broad areas of signal, image, and video processing \cite{Vai93Book, VetKov95Book, StrNgu96}. 
Fig.~\ref{fig::multichannel}(a) shows a $C$-channel $D$-fold subsampled filter bank (FB).  
Here, a FB is considered to achieve PR if it reconstructs an exact though possibly delayed replica $x[n-n_0]$ of the input $x[n]$, that is,  \mbox{$\hat x[n] = x[n-n_0]$} for some integer delay $n_0$. 
Filter banks can be categorized as: (\emph{i}) critically sampled or maximally decimated, i.e., when the downsampling factor $D$ equals the number of channels $C$; or (\emph{ii}) \emph{oversampled}, when there are more channels than the downsampling factor, i.e., $C>D$.  

For critically sampled FBs,  
the PR requirement is typically in conflict with other desirable design specifications. 
In the oversampled case, however, with a given set of analysis filters, an infinite number of PR synthesis filters exist  
\cite{BolHlaFei98}. 
The main advantage of oversampled FBs are the added degrees of design freedom  gained from this redundancy, which have been exploited for the reduction of quantization noise in digital communication systems \cite{bolcskei1997, BolHla0101}, for improved equalization and precoding in data communication \cite{ScaGiaBar9907, LabChiKie0512a}, and in image transmission \cite{KovDraGoy0206} and image coding \cite{GanMa0410}.

\vspace{-0.1in}
\subsection{The role of FBs in multi-channel systems}

Multi-channel data acquisition/sampling arises in various sensing, imaging, and data processing modalities including data communication/storage applications, remote sensing/imaging, and medical imaging systems such as magnetic resonance imaging (MRI) \cite{RefWorks:491}. The continuous-time model for a $C$-channel sampling (data acquisition) and reconstruction system is illustrated in Fig.~\ref{fig::multichannel}(b). 
The channel outputs are sampled prior to digital processing, say, with a uniform sampling rate $T$.  The objective is to perfectly reconstruct the input signal from the sampled output signals $y_i[n] = y_i(nT)$, i.e., to design the signal \emph{reconstruction} mechanism in Fig.~\ref{fig::multichannel}(b), so that $\hat{x}(t) = x(t).$ 

If the channel characteristics are known, the problem reduces to a well-studied problem in sampling theory, widely knowns as Papoulis generalized sampling \cite{Pap7711, Unser00} 
and its
generalizations to the oversampled case 
\cite{Unser00, MarksBook91}
and to multi-dimensional (MD) sampling \cite{PetMid62} of MD signals \cite{Ize0506}.   
%
Examples of MD multi-channel systems include sensor arrays with sampling in space and time, and multi-channel (parallel) MRI \cite{RefWorks:491} with spatio-temporal sampling \cite{RefWorks:441, ShaDerFarBre10mrm}.  

The analysis simplifies by a fully discrete-time formulation. Under mild conditions \cite{VenBre0312}, we can convert the continuous-time channel model in Fig.~\ref{fig::multichannel}(b) into the discrete-time model (FB structure) in Fig.~\ref{fig::multichannel}(a), where $x[n]$ and $\hat{x}[n]$ represent samples of $x(t)$ and $\hat{x}(t)$ taken at a sufficiently high rate, and the $D$-fold subsampling models the sub-Nyquist sampling of the channels in Fig.~\ref{fig::multichannel}(b). Although all channels in Fig.~\ref{fig::multichannel}(a) have identical uniform subsampling, this setup is fairly general and subsumes periodic nonuniform subsampling \cite{VenBre0312a}.
It follows that FBs represent equivalent discrete-time models for a wide class of multi-channel data acquisition/sampling systems.

\vspace{-0.1in}
\subsection{Generic FBs: motivation for short synthesis filters} 

In most FB-related problems, it is assumed that one has the luxury to almost freely manipulate the analysis filters, and the goal is to design the \emph{entire} FB with desirable properties such as robustness to noise or erasures,   frequency selectivity, etc. \cite{BolHlaFei98, BolHla0101, KovDraGoy0206}.  
However, considering FBs as models for multi-channel systems, often little or no control can be exerted over the analysis filters since their characteristics are dictated by the underlying physics.  
This is the case in important sensing, imaging, and image processing applications. Examples of such ``sensor filters'' include spatial sensitivity of receivers in multi-channel MRI \cite{RefWorks:491, Griswold:2002km, ShaBre10ismrm1, ShaBre1103},  
or blurring kernels associated with remote imaging applications \cite{MolNunCor01, Blahut}.  
Furthermore, since the sensor filters result from complex physical processes,  a pathological case would be extremely unlikely; in other words, the corresponding analysis filters are ``generic'' (further described in Section \ref{sec::preliminary_notations}).  
Therefore, in this work, we focus on scenarios wherein the set of analysis filters, also called the \emph{analysis bank}, is \emph{fixed and generic}. 
Such a \emph{generic analysis} covers most practical applications of multi-channel sensing and imaging.  
It has the added advantage of revealing the properties that are inherited from the FB \emph{structure} rather than specific values of the filter taps.   

Another important aspect of FBs is the length of the filters in the synthesis bank: infinite or finite impulse response (IIR/FIR). 
Given the analysis bank, the PR synthesis bank is typically designed to equal (or closely approximate) the so-called para-pseudoinverse or dual-frame synthesis bank \cite{BolHlaFei98, CveVet9805}, 
to minimize the reconstruction noise gain.  
%
Unless the analysis bank meets stringent condition \cite{CveVet9805, Bol99}, which almost surely do not hold for generic FBs, the dual-frame synthesis is IIR.  
%
Even if the exact PR condition is relaxed to approximate PR, accurate approximations of the dual-frame synthesis bank requires long synthesis filters \cite{Str99,Str01}.\footnote{A ``snug'' (almost tight) analysis frame is an exception \cite{BolHlaFei98, Str01}, which does not occur for generic FBs.} 
An alternative to such approaches is to search for the ``best'' FIR synthesis bank that achieves PR --- and simultaneously satisfies additional optimality criteria \cite{Tan06, GauDuvPes0910}.   
In particular, short FIR PR synthesis banks have important advantages including the following:
\begin{enumerate}
\item[--] 
Short PR FB reconstruction is computationally very efficient, as compared to the IIR dual-frame synthesis bank or alternative non-FB reconstructions such as the least-squares solution. 
\item[--] 
A short FIR synthesis bank implies a low-dimensional search space for the synthesis FB design algorithm, hence reducing the computational complexity and, more importantly, improving the optimized design. This is especially significant for practical design techniques that employ additional desired criteria (besides PR) in  a non-convex optimization scheme (e.g., \cite{GauDuvPes0910}).
\item[--] 
In certain applications, such as data communications or storage, the multi-channel data is corrupted by ``impulsive noise'' \cite{GabRioDuh01, LabChiKie0512}, 
which can also model channel erasures \cite{KovDraGoy0206}, 
or dead pixels in a sensor.  
The noise-optimal dual-frame synthesis bank (generically IIR) or non-FB reconstruction using the maximum-likelihood (least squares) solution would corrupt the \emph{entire} reconstructed signal. 
Instead, short FIR synthesis can provide ``good'' reconstruction in such scenarios,  achieving PR everywhere except in the neighborhood of the noise spikes \cite{HarBre99b}. 
\end{enumerate}
These advantages and the prevalence of the generic FB scenario in multi-channel systems motivate the work in this paper.

\vspace{-0.1in}
\subsection{Present Work}

This work addresses the feasibility of short FIR synthesis for PR  (allowing for reconstruction delay --- \emph{delayed} PR) in generic complex-valued FIR one-dimensional (1D) FBs.
Specifically, we address the following two questions for a $C$-channel FIR generic analysis FB with $D$-fold subsampling and analysis filter length $m_h$:

\noindent \textbf{[Q.1]} \emph{Feasibility of FIR Synthesis}: What are the conditions for existence of an FIR synthesis bank that achieves PR or delayed PR?\\
\noindent \textbf{[Q.2]} \emph{Feasibility of Short Synthesis and Effect of Oversampling}: what is the \emph{minimum feasible filter length} $m_v^{\ast}$ among all PR (or delayed PR) synthesis banks ---  
and how does it depend on the oversampling factor $C/D$?

With $D=1$, the FB is \emph{nonsubsampled} \cite{CveVet9805} and the reconstruction problem is equivalent to multi-channel deconvolution. For this case, Q.1 has been extensively studied in the theory of polynomials \cite{BerYge93,BerPat9004}. Further, Q.2 has been addressed in previous work by Harikumar and Bresler in 1D (with $C\geq 2$) \cite{HarBre98a} and 2D (with $C\geq 3$) \cite{HarBre99b}. In the 1D case with generic channels, they provide prescriptions for deconvolver lengths that are both necessary and sufficient for PR \cite{HarBre98a}.

For the subsampled ($D>1$) case, although the conditions for PR for a given critically sampled and oversampled FB have been known for many years \cite{Vai93Book, CveVet9805}, Q.1 has only been recently answered by Law et al.~\cite{LawFosDo0911}, in fact for the more general case of MD signals. (We discuss their relevant result in Section \ref{sec::bounds_n_relations}.) However, to the best of our knowledge, Q.2 is an open problem.

The significance of Q.2 is perhaps best appreciated in the context of \emph{blind} signal PR, i.e., the
perfect inversion of \emph{subsampled} multi-channel systems (i.e., the analysis FB)  by \emph{identification} of a PR synthesis bank without any prior knowledge of the channels. 
For this class of problems only the nonsubsampled case, i.e., the problem of blind multi-channel deconvolution, is fully studied (see \cite {TonPer9810,AbeQiuHua9708} and references therein). 
%
In subsampled systems little rigorous analysis is available of \emph{both} necessary and sufficient conditions for blind PR (for relevant work in 2D see \cite{ElaFeu9712,Yag03,SroCriFlu0709} and references therein).  
In such problems, where an FIR synthesis bank is to be identified, the length (support size) of the synthesis filters is a fundamental issue as it dictates the dimensionality of the unknown parameter space.  
In fact, given \emph{limited available data}, as the length allocated for the synthesis filters increases (more unknowns to solve for), the inverse problem of estimating the synthesis filters becomes progressively more difficult and/or ill-posed.  
(This is compounded by the concomitant increase in  computational cost  with more unknowns.)  
Therefore, it is important to constrain the FB length. On the other hand, reducing the dimensionality  too much would make PR infeasible. Consequently, knowing where this \emph{phase transition} between PR feasibility/infeasibility occurs becomes critical.  
Having the answer to Q.2, one would be able to (i) find the minimally required dimensionality of the parameter space to enable PR; and (ii) analyze the fundamental trade-offs, such as the oversampling factor needed to guarantee a ``feasible'' (low dimensional) search space.

In this work, we address the above-raised two questions for generic FIR analysis banks that are at least 2-fold oversampled; this subsumes most practical cases of oversampled FBs. We show that with such oversampling factors, PR is almost surely feasible with a synthesis bank that consists of filters as short or shorter than the analysis filters. Furthermore, we show that  
the required length for the synthesis filters has an inverse relationship to the oversampling factor.

Our  results indicate that satisfying the PR condition per se is quite easy --- even a random choice will do --- if the filter lengths satisfy certain conditions. 
Hence,  
we can guarantee feasibility of exact PR in the design of a FB by prescribing  the analysis/synthesis filter lengths. 
This implies that the degrees of freedom in the design process can be mostly  driven by other desired criteria, e.g., reconstruction noise gain \cite{bolcskei1997}, frequency selectivity, time/frequency localization, subband attenuation \cite{GanMa0303}, or coding gain \cite{Lab0510} --- all while guaranteeing PR.   
For the problem of designing a PR synthesis bank given the analysis bank, our results provide an alternative to the expensive  exhaustive search for the synthesis filter length \cite{GauDuvPes0910}.   

The paper is organized as follows. Section \ref{sec::preliminary} contains basic definitions and notations. In Sections \ref{sec::SuffCond} and \ref{sec::NecCond}, respectively, we present necessary and sufficient requirements 
on the minimal filter length for the PR synthesis bank.   
Section \ref{sec::bounds_n_relations} uses these results  to address Questions Q.1 and Q.2. In Section \ref{sec::results_filterlength}, we provide numerical verification of the theoretical results; further, we study the feasibility of near-PR using synthesis filter lengths below those prescribed by our propositions. Finally, Section \ref{sec::disc_and_conc} summarizes the results and concludes the paper.

\section{Preliminaries}\label{sec::preliminary}

\subsection{Notations and Setup}\label{sec::preliminary_notations}

For $a\in\mbb{R}$, $\lceil a\rceil$ (respectively, $\lfloor a\rfloor$) denotes the smallest (respectively, largest) integer larger (respectively, smaller) than or equal to $a$. 
Column vectors and matrices are denoted by lowercase and uppercase letters, respectively. The elements of a vector $a$ are indexed as $a[i]$, $0\leq i \leq m_a-1$; similarly, the elements of a matrix are denoted as $A[i,j]$, with upper-left element $A[0,0]$.  
Signals and column vectors are used interchangeably. 
 The length of a signal or vector $s$ is denoted by $m_s$. 
%
For a matrix $A$, its transpose and Hermitian are denoted by $A^{\mathrm{T}}$ and $A^{\mathrm{H}}$, respectively; $\Range(A)$ is its range (column) space and $A^{\dagger}$ is its Moore-Penrose pseudoinverse. 
%
%
The notation $\mathrm{vec}[A]$ denotes the vector obtained by concatenating columns of $A$ in lexicographical order. Similarly, concatenation of a sequence of vectors $\{a_i\}_{\scsc i=1}^{\scsc N}$ into a single vector is denoted by $\mathrm{vec}\big[\{a_i\}_{\scsc i=1}^{\scsc N}\big]$.   

The shifted unit pulse $\vv\delta_{m}(N)$ is defined as the $(m+1)$-th column of the $N\times N$ identity matrix, $I_N$. 
In most cases, the $N$ argument in $ \vv\delta_{m}(N)$  can be inferred from the context and is dropped for notational brevity. 
%
The convolution of $s[n]$ and $h[n]$ is denoted by $(s\ast h)[n]$ and is equivalently written in vector form as $\cconv_{m_s}\{h\} \vv s$, where $\cconv_{m_s}\{h\}$  is the matrix representation of the convolution operator, which is  a Toeplitz matrix of size of $(m_h  + m_{s} - 1 ) \times m_s$.
%
%
Finally, we define the ``stack'' of all convolution matrices corresponding to the analysis channels as follows:
\beqn
\Cbar
\Big[\left\{ {h_{\,i} } \right\}_{\scsc  i = 1}^{\scsc C} \Big]_{m_s} =\Big[ \cconv_{m_s}\{h_1\}\;\; \cconv_{m_s}\{h_2\} \;\; \dots \, \;  \cconv_{m_s}\{h_{\scsc C}\} \Big].
\eeqn

Consider the standard $C$-channel filter bank (FB) structure with $D$-fold subsampling shown in Fig.~\ref{fig::multichannel}(a).  We focus on \emph{oversampled} FB, i.e., where the \emph{oversampling factor} $C/D > 1$.  
The transfer functions of the analysis and synthesis filters are denoted by $H_i(z)$ and $V_i(z)$, respectively, and their corresponding impulse responses, assumed to be FIR, by $h_i[n]$ and  $v_i[n]$. For brevity, we refer to the set of analysis (respectively, synthesis) filters as the analysis (respectively, synthesis) bank.  
It is assumed that the support of all filters in the analysis (respectively, synthesis) bank is the same and --- without loss of generality --- is right-sided (that is, the filters are causal). Consequently, all analysis (respectively, synthesis) filters have equal length, denoted by $m_h$ (respectively, $m_v$).\footnote{This is not a limiting assumption as one can take the length of a set of filters $\{f_i[n]\}_{i=1}^{\scsc C}$ to be $m_f = \max m_{f_i}$.}  For example, the support for $h_i[n]$ is \mbox{$0\leq n \leq m_h-1$}.
We therefore have the following expressions for the $z$-transforms of the filters: $H_i(z) = \sum_{n=0}^{m_h-1} h_i[n] z^{-n}$ and $V_i(z) = \sum_{n=0}^{m_v-1} v_i[n] z^{-n}$.   

Most of the theoretical work here involves study of ``generic'' properties of vectors and matrices. 
We use the same definition for a property to hold ``generically'' as in previous works \cite{HarBre99b, HarBre98a}: 
If a property $\mathcal{P}$ of a vector $\vv a\in \mbb{R}^n$ fails to hold only on a closed set of measure zero that is nowhere dense\footnote{A set $S$ is dense in $\mbb{R}$ if for all $x\in\mbb{R}$, any neighborhood of $x$ contains at least one point from $S$. For example, the rational numbers are dense in $\mbb{R}$. A complete definition can be found in \cite{Roy89}.} in $\mbb{R}^n$, we say that the property holds for \emph{generic} $\vv a$, or equivalently, $\mathcal{P}$ holds \emph{generically}. 
As a result, $\mathcal{P}$ will hold with probability 1 (short form: ``w.p.1'') when the elements of the vector $\vv a$ are drawn independently from a probability distribution that admits a probability density function\footnote{The probability distribution should be absolutely continuous with respect to Lebesgue measure.}
Furthermore, property $\mathcal{P}$ is \emph{robust}, in the sense that $\mathcal{P}$ continues to hold for any sufficiently small perturbation of such a randomly generated $\vv a$. 
A more mathematically rigorous definition \cite{LawFosDo0911}, or an alternative but equivalent notion of a generic property \cite{PaiHavBov98} can be found elsewhere.

\subsection{PR in Polyphase Domain}\label{sec::preliminary_polyph}

The \emph{polyphase decomposition} \cite{Vai93Book, VetKov95Book} of the analysis filters $H_i(z)$ is given by
\beqn
H_i(z) = \sum_{p=0}^{D-1} z^{p} \: \H_{i, p}(z^{D}),
\textrm{\: where\quad}
\H_{i, p}(z) = \sum_{n=-\infty}^{\infty} h_i [nD - p] z^{-n}
\label{eq::def_HzEnteriesPolyph}
\eeqn
is the $p$-th polyphase component ($p=0,\dots,D-1$) of the $i$-th analysis filter.  
The polyphase decomposition of the synthesis filters is similar, but with opposite signs for the index $p$:
\beqn
V_i(z) = \sum_{p=0}^{D-1} z^{-p} \: \V_{i, p}(z^{D}), \quad 
\V_{i, p}(z) = \sum_{n=-\infty}^{\infty} v_i [nD + p] z^{-n}. 
\eeqn
%
 There is a corresponding time-domain representation on both analysis and synthesis sides that, for each analysis/synthesis channel,  involves partitioning of the filters taps into $D$ subsequences --- based on congruency of their indices modulo $D$.

Define the impulse response  corresponding to $\H_{i, p}(z)$ as: $h_{i,p}[n] = h_i [nD - p]$; and denote its length by $m_{h;p}$, which is equal to  $m_{h;p} = \ceilp{\frac{m_h - p}{D}}$.  
Similarly, the length of the impulse response $v_{i,p}[n] = v_i [nD + p]$ corresponding to $\V_{i, p}(z)$ is denoted by $m_{v;p}$ and is equal to $m_{v;p} = \big\lceil \frac{m_v - p}{D} \big\rceil$.   
The following properties, for $\alpha = h$ and $\alpha = v$, are easy consequences of these definitions:
\beqn
\textrm{(a)~} \sum_{p=0}^{D-1} m_{\alpha;p} = m_{\alpha}   \quad\:\:\: \textrm{(b)~}      \Big\lfloor \frac{m_{\alpha}}{D} \Big\rfloor \leq m_{\alpha;p} \leq \Big\lceil\frac{m_{\alpha}}{D} \Big\rceil  \quad\:\:\: \textrm{(c)~}  m_{\alpha;\po} = \Big\lfloor \frac{m_{\alpha}}{D}\Big\rfloor \textrm{ for some } 0\leq\po\leq D-1.
\label{eq::polyphlength_properties}
\eeqn
%

Based on the theory of filter banks \cite{Vai93Book, VetKov95Book}, the \emph{polyphase-domain} condition for PR with an output delay of $n_0 = m_0D$ for all inputs $x[n]$ is as follows:
\begin{eqnarray}
\Big[ {\begin{array}{*{20}c} \!  {\V_{1,p} (z)},  {\V_{2,p} (z)}, \,\hdots, {\V_{C,p} (z)} \!\!\! \end{array}} \Big] \; 
 \sH(z) = z^{ - m_0 }\:\vv\delta_{p}^{\,\mathrm{T}}(D)
 \quad \qquad \forall z\in\mbb{C,} \qquad p=0, \hdots, D-1,
\label{eq::polyphcond_z}
\end{eqnarray}
where the so-called \emph{analysis polyphase matrix} $\sH(z)$, which is a $C \times D$ (Laurent) polynomial matrix, has entries $\sH_{i,j}(z) = \H_{i+1,j} (z), \; i=0, \hdots, C-1, \; j=0, \hdots, D-1$. 
Collecting all $D$ polyphase conditions in \eqref{eq::polyphcond_z} into a single equivalent PR condition yields
\begin{eqnarray}
 \sV(z) \; \sH(z)  = z^{ - m_0 }\:I_{D}  \qquad \quad \forall z\in\mbb{C},
 \label{eq::polyphPRcond_matrixversion} 
\end{eqnarray}
where the $D \times C$ matrix $\sV(z)$ with entries 
$\sV_{i,j}(z) = \V_{j+1,i} (z), \; i=0, \hdots, D-1, \; j=0, \hdots, C-1$, 
is referred to as the \emph{synthesis polyphase matrix}. For the case of zero delay,  \eqref{eq::polyphPRcond_matrixversion} states the PR is achieved when the synthesis polyphase matrix is a left inverse of the analysis polyphase matrix, for all $z\in\mbb{C}$. 
The sampling- (time-) domain counterpart of \eqref{eq::polyphcond_z} can be written in the following form 
\beqn
& \underbrace {\left[ {\begin{array}{*{20}c}
   \Cbar { [\left\{ {h_{i,0} } \right\}_{\scsc  i = 1}^{\scsc C} ]} _{m_{v;p}} \\
   \Cbar \,[\left\{ {h_{i,1} } \right\}_{\scsc i = 1}^{\scsc C} ] _{m_{v;p}} \\
    \vdots   \\
   \Cbar \,[\big\{ {h_{i,(D-1)} }\big\}_{\scsc i = 1}^{\scsc C} ] _{m_{v;p}}  \\
 \end{array} } \right]} \mathrm{vec}\Big[\{v_{i,p}\}_{\scsc i=1}^{\scsc C}\Big] = & \vv\delta_{\kappa(p,m_0 D)} \qquad \nonumber \\
& \bH_p \qquad \qquad \qquad &  p=0,\dots,(D-1),
\label{eq::polyphcond_time}
\eeqn
where $\bH_p$ is the sampling-domain analysis polyphase matrix and is of size $(m_{h} + Dm_{v;p} - D)\times (C \: m_{v;p})$.\footnote{ In \eqref{eq::polyphcond_time}, each of the matrices $\Cbar { [\left\{ {h_{i,k} } \right\}_{\scsc  i = 1}^{\scsc C} ]} _{m_{v;p}} $ for $k=0,\dots,(D-1)$ is of size  \mbox{$(m_{h;k} + m_{v;p} - 1)\times (C \: m_{v;p})$}. Therefore, using  $\sum_{p=0}^{D-1} m_{h;p} = m_h$, it is seen that the matrix $\bH_p$ is of size $(m_{h} + Dm_{v;p} - D)\times (C \: m_{v;p}).$
}
Assuming $0\leq m_0 \leq \lfloor \frac{m_h}{D} \rfloor+\lfloor \frac{m_v}{D} \rfloor-2$, the right-hand side is a shifted unit pulse with the amount of shift $\kappa(p,m_0 D)$ given by\footnote{For \eqref{eq::kappa_def} to hold we need $0\leq m_0\leq m_{h;k} + m_{v;p} - 1$ for $0\leq k,p\leq D-1$, which implies the condition on $m_0$ given above.}
\beqn
\kappa(p,m_0 D)  = 
\left\{\begin{array}{cl} 
m_0 \quad &  p=0	 \\ 
m_0 + \sum_{k=0}^{p-1} (m_{h;k} + m_{v;p} - 1)  \quad &   p=1,\dots,(D-1).
\end{array} \right. 
\label{eq::kappa_def}
\eeqn

The abovementioned PR conditions correspond to cases where the delay allowed in PR is a multiple of the subsampling factor, i.e., $n_0= m_0 D$. Nevertheless, the PR condition in \eqref{eq::polyphPRcond_matrixversion} can be extended to the general delayed PR with a delay of $n_0 = m_0 D + r_0$, $0\leq r_0\leq D-1,$ as follows (cf.~\cite{Vai93Book},  Ch.~5.6): 
\begin{eqnarray}
\sV(z) \sH(z) =  z^{ - m_0 }\Bigg[\,\begin{matrix} O  & z^{-1}I_{r_0} \\ I_{D-{r_0}} & \acute O \end{matrix}\,\Bigg],
 \label{eq::polyphPRcond_genDelay}
 \end{eqnarray}
where, in the $D\times D$ matrix on the right-hand side, $O$ and $\acute O$ denote zero matrices of appropriate size. 
The corresponding sampling-domain condition will 
differ from \eqref{eq::polyphcond_time} only in the location of the 1 in the delay vector. In short, assuming\footnote{This assumption can be relaxed to allow for larger $n_0$ by appropriate zero-padding of the analysis filter taps.}  $0\leq \lceil\frac{n_0}{D}\rceil \leq \lfloor \frac{m_h}{D} \rfloor+\lfloor \frac{m_v}{D} \rfloor-2$, the PR condition with $n_0 = m_0D +r_0$ delay is
\beqn
\makebox[3in][r]{$\displaystyle \bH_p \mathrm{vec}\Big[\{v_{i,p}\}_{\scsc i=1}^{\scsc C}\Big] =  \vv\delta_{\kappa(p,n_0)}$} \quad \qquad p=0,\dots,(D-1),
\label{eq::polyphcond_time_genDelay}
\eeqn
where  $\vv\delta_{\kappa(p,n_0)}$ is the inverse $z$-transform of the $p$-th row of the right-hand side of \eqref{eq::polyphPRcond_genDelay}.  
The assumption on the range for $n_0$   is needed for \eqref{eq::polyphPRcond_matrixversion} to be feasible; however, the present formulation can account for delays outside this range by proper zero-padding of analysis/synthesis impulse responses. 
The general closed-form expression for $\kappa(p,n_0)$, given in \eqref{eq::kappa_def} for the special case of $n_0=m_0D$, is somewhat complicated and of no significance in this paper; hence, it is skipped here.

To illustrate the structure of the sampling-domain analysis polyphase matrix $\bH_p$ given in \eqref{eq::polyphcond_time}, consider for example a $C=6$ channel FB with $D=3$, $m_h=7$, and $m_v=6$.  The corresponding $\bH_2$ is shown in Fig.~\ref{fig::Example_Hp_Matrix}. 
In general, the structure of $\bH_p$ consists of Toeplitz  (rectangular) \emph{blocks} of the form $\cconv_{m_{v;p}}\{h_{c_\circ,\ell}\}$ with $C$ \emph{block-columns} ($1$$\leq$$c_\circ$$\leq$$C$) and $D$ \emph{block-rows} ($0$$\leq$$\ell$$\leq$$D-1$). 
The zeros in the Toeplitz blocks are referred to as \emph{structural zeros} --- they are underlined to distinguish them from \emph{assigned zeros} in our matrix constructions in the following sections. On occasion, we need to refer to an \emph{indeterminate} zero, i.e., one that can be either assigned or  structural, for which we use the notation $\dcz$. 
%

When $m_h<D$ or $m_v<D$, the system of equations in \eqref{eq::polyphcond_time_genDelay} should be interpreted with some care:  equations that correspond to $m_{v;p}=0,$  and    blocks in $\bH_p$ corresponding to polyphase components in $\{h_i\}$ that do not exist, should all be removed. Because of these complications and considering that the case $m_h<D$ is of limited practical interest,  we assume $m_h\geq D$ throughout. 


Finally, given the analysis filters, the following sampling-domain necessary and sufficient condition for existence of a PR synthesis bank follows immediately from \eqref{eq::polyphcond_time_genDelay}. 

\newtheorem{lemma}{Lemma}
\begin{lemma}\label{thm::PRCond}  
For a $C$-channel FB with given   FIR analysis filters $\{ h_i[n]\}_{i=1}^{C}$,  a set of length-$m_v$  synthesis filters  achieving PR with  delay $n_0$, $0\leq \lceil\frac{n_0}{D}\rceil \leq \lfloor \frac{m_h}{D} \rfloor+\lfloor \frac{m_v}{D} \rfloor-2$, exists if and only if  
$\displaystyle \vv \delta_{\kappa(p,n_0)} \in \Range (\bH_p),$  for all $~p=0,\dots,D-1.$ 
\end{lemma}

\section{Minimum Length for PR Synthesis Filters: Generic Sufficient Condition}\label{sec::SuffCond}

In this section we aim to answer Question Q.2 raised in Section \ref{sec::intro}. Specifically, for a length-$m_h$ generic analysis FB with $C$-channels and $D$-fold subsampling, we   propose a functional $m_v^{\st S}(C,D,m_h):\mbb{N}^3\rightarrow \mbb{N}$ such that for all integers $m_v\geq m_v^{\st S}(C,D,m_h)$ there   exists a PR synthesis bank with filter length $m_v$.   
%


\vspace{-0.15in}
\subsection{Statement of the result}\label{sec::SuffCond_statement}


\begin{definition}
Denote by $m^{\st S}_v(C,D,m_h)$ the minimal value of $m_v\in\mathbb{N}$ that satisfies: 
\beqn
\frac{C}{D} \geq 1 + \frac{1}{D}\sum_{p=0}^{D-1} \Bigg\lceil\frac{m_{h;p}  - 1}{\lfloor m_v/D \rfloor}\Bigg\rceil,  
\label{eq::suffcond_condition}
\eeqn
where $m_{h;p} = \Big\lceil \frac{m_h - p}{D} \Big\rceil$.
We refer to $m_v^{\st S}$ as the sufficient synthesis filter length, or in short the \emph{sufficient length}.
\end{definition}
%
\vspace{-9pt}
It is easy to show that  \eqref{eq::suffcond_condition} is satisfied for any $m_v\geq m^{\st S}_v(C,D,m_h)$. Therefore, the set of $m_v\in\mbb{N}$ satisfying \eqref{eq::suffcond_condition} is a right-sided interval, i.e., all integers in: $\Big[m_v^{\st S}(C,D,m_h), \infty\Big)$.  
%
The following lemma shows that for $m^{\st S}_v$ to be finite, at least two-fold oversampling is needed (assuming $m_h\geq 2D$,  which is the case in most practical scenarios). The proof is provided in Appendix \ref{sec::app_proof_mvstar}.
\vspace{-6pt}
\newtheorem{lemma_suff_required}[lemma]{Lemma}
\begin{lemma_suff_required}\label{lem::FullRankGenericSuffRequired}  
Suppose $m_h\geq 2D$. Then, if $C/D < 2$, \eqref{eq::suffcond_condition} is not satisfied for any finite $m_v$. 
\end{lemma_suff_required}
\vspace{-6pt}

In the following proposition, the main result of this section, we consider generic FIR analysis banks that are at least 2-fold oversampled, which, as stated in Lemma \ref{lem::FullRankGenericSuffRequired}, is a requirement for the sufficient-length condition in \eqref{eq::suffcond_condition} to be feasible for FIR synthesis banks. We will show that PR is generically feasible with a synthesis bank that consists of filters with lengths $m_v\geq m_v^{\st S}(C,D,m_h)$.

\vspace{-6pt}
\newtheorem{theorem}{Proposition}
\begin{theorem}\label{thm::FullRankGenericSuff} 
\textbf{\textsc{(}sufficient length\textsc{)}}~
A $D$-fold subsampled $C$-channel length-$m_h$ FIR analysis FB with $C/D \geq 2$ is generically invertible, i.e., the FB admits PR with any delay $n_0$, $0\leq \lceil\frac{n_0}{D}\rceil \leq \lfloor \frac{m_h}{D} \rfloor+\lfloor \frac{m_v}{D} \rfloor-2$, by a length-$m_v$ synthesis FB if $m_v\geq m_v^{\st S}(C,D,m_h)$. 
\end{theorem}

\begin{figure}[t!]
\centering
\beqn
\bH_{2} = \left[
\scriptsize 
\begin{tabular}[c]{ c c | c c | c c | c c | c c | c c }
 $h_1[0]$ & $\0$          & $h_2[0]$ & $\0$           & $h_3[0]$ & $\0$           & $h_4[0]$ & $\0$  & $h_5[0]$ & $\0$ & $h_6[0]$ & $\0$\\
 $h_1[3]$ & $h_1[0]$  & $h_2[3]$ & $h_2[0]$  & $h_3[3]$ & $h_3[0]$  & $h_4[3]$ & $h_4[0]$  & $h_5[3]$ & $h_5[0]$ & $h_6[3]$ & $h_6[0]$\\
 $h_1[6]$ & $h_1[3]$  & $h_2[6]$ & $h_2[3]$  & $h_3[6]$ & $h_3[3]$  & $h_4[6]$ & $h_4[3]$  & $h_5[6]$ & $h_5[3]$   & $h_6[6]$ & $h_6[3]$ \\
 $\0$         & $h_1[6]$  & $\0$          & $h_2[6]$  & $\0$         & $h_3[6]$  & $\0$          & $h_4[6]$  & $\0$         & $h_5[6]$ & $\0$         & $h_6[6]$\\[1pt]
\hdashline
 $h_1[1]$ &  $\0$ & $h_2[1]$ &  $\0$ & $h_3[1]$ & $\0$  & $h_4[1]$ &  $\0$   & $h_5[1]$ & $\0$  & $h_6[1]$ & $\0$\\
 $h_1[4]$ & $h_1[1]$  & $h_2[4]$ & $h_2[1]$  & $h_3[4]$ & $h_3[1]$  & $h_4[4]$ & $h_4[1]$  & $h_5[4]$ & $h_5[1]$  & $h_6[4]$ & $h_6[1]$\\
 $\0$ & $h_1[4]$  &  $\0$ & $h_2[4]$  &  $\0$ & $h_3[4]$  &  $\0$ & $h_4[4]$  &  $\0$ & $h_5[4]$  &  $\0$ & $h_6[4]$\\[1pt]
\hdashline
 $h_1[2]$ &  $\0$  & $h_2[2]$ &  $\0$  & $h_3[2]$ &  $\0$  & $h_4[2]$ &  $\0$    & $h_5[2]$ &  $\0$ &  $h_6[2]$ &  $\0$\\
 $h_1[5]$ & $h_1[2]$  & $h_2[5]$ & $h_2[2]$  & $h_3[5]$ & $h_3[2]$  & $h_4[5]$ & $h_4[2]$   & $h_5[5]$ & $h_5[2]$ &  $h_6[5]$ & $h_6[2]$ \\
 $\0$ & $h_1[5]$  &  $\0$ & $h_2[5]$  &  $\0$ & $h_3[5]$  &  $\0$ & $h_4[5]$  &  $\0$ & $h_5[5]$ &  $\0$ & $h_6[5]$ \\
\end{tabular}
\right]
\nonumber
\eeqn
\vspace{-0.1in}
\caption{\small 
Structure of the sampling-domain analysis polyphase matrix $\bH_{2}$ (for $p=2$) of size $10\times 12$ corresponding to a 6-channel analysis FB with $3$-fold subsampling ($D=3$) and a filter length of $m_h=7$ (i.e., $m_{h;0}=3, m_{h;1}=m_{h;2}=2$). The synthesis filter length is $m_v=6$ (i.e.,  $m_{v;2}=2$). The structural zeros are underlined. 
}
\label{fig::Example_Hp_Matrix}
\vspace{-0.2in}
\end{figure}
%
%
\begin{figure}[t!]
\centering
\beqn
\bH_{2} = \left[
\tiny
\begin{tabular}[c]{ c c | c c | c c | c c | c c | c c}
     \f{1}  &   \0  &   0  &   \0  &   0  &   \0  &   0  &   \0  &   0  &   \0  &   0  &   \0 \\
     0  &   \f{1}  &   0  &   0  &   0  &   0  &   0  &   0  &   0  &   0  &   0  &   0 \\
     0  &   0  &   \f{1}  &   0  &   0  &   0  &   0  &   0  &   0  &   0  &  0  &  0 \\ 
     \0  &   0  &   \0  &   \f{1}  &   \0  &   0  &   \0  &   0  &   \0  &   0  &   \0  &   0 \\[1pt]
     \hdashline
     0  &   \0  &   0  &   \0  &   \f{1}  &   \0  &   0  &   \0  &   0  &   \0  &   0  &  \0 \\
     0  &   0  &   0  &   0  &   0  &   \f{1}  &   \f{1}  &   0  &   0  &   0  &   0  &   0 \\
     \0  &   0  &   \0  &   0  &   \0  &   0  &   \0  &   \f{1}  &   \0  &   0  &   \0  &   0 \\[1pt] 
     \hdashline
          0  &   \0  &   0  &   \0  &   0  &   \0  &   0  &   \0  &   \f{1}  &  \0  &   0  &   \0 \\
     0  &   0  &   0  &   0  &   0  &   0  &   0  &   0  &   0  &   \f{1}  &   \f{1}  &   0 \\
     \0  &   0  &   \0  &   0  &   \0  &   0  &   \0  &   0  &   \0  &   0  &   \0  &   \f{1} \\
\end{tabular}
\right]
\nonumber
\eeqn
\vspace{-0.1in}
\caption{\small  Analysis polyphase matrix $\bH_p$ (for $p=2$), with the same structure as in Fig.~\ref{fig::Example_Hp_Matrix}, computed using Algorithm \ref{alg::one} for a 6-channel analysis FB with $D=3$ and $m_h=7$. 
The synthesis filter length is taken to be the sufficient length: $m_v = m_v^{\st S}(C,D,m_h) = 6$. The assigned ones are boxed and the structural zeros are underlined. 
}
\label{fig::Example_Suff_proof}
\vspace{-0.1in}
\end{figure}
%
%
\begin{figure}[t!]
\centering
\subfloat{
\hspace{-24pt}
\tiny
\begin{tabular}[c]{ c c | c c }
    $\:\:\:\quad\vdots$ & $\vdots$&  $\vdots$& $\vdots$\\
    \:\:\:\dots\:\f{1}  &   \dcz  &   \dcz  &   \dcz  \\
    \dots\:\:  \dcz  &   \f{1}  &   \dcz  &   \dcz  \\
   \dots\:\:   \dcz  &   \dcz  &   \f{1}   & \dcz   \\
   \dots\:\:   \dcz  &   \dcz  &   \dcz  &   \f{1}  \\
\end{tabular}
\hspace{8pt}
}
\subfloat{
\tiny
\begin{tabular}[c]{ c c | c c c }
    $\:\:\:\quad\vdots$ & $\vdots$&  $\vdots$& $\vdots$& $\vdots$\\
    \:\:\:\dots\: \f{1}  &   \dcz  &   \dcz  &   \dcz &   \dcz\\
     \dots\:\: \dcz  &   \f{1}  &   \f{1}  &   \dcz  &   \dcz\\
     \dots\:\: \dcz  &   \dcz  &   \0  &   \f{1}  &   \dcz\\
     \dots\:\: \dcz  &   \dcz  &   \0  &   \0  &   \f{1}\\
\end{tabular} 
\hspace{6pt}
}
\subfloat{
\tiny
\begin{tabular}[c]{ c c c | c c c }
    $\:\:\:\quad\vdots$ & $\vdots$&  $\vdots$& $\vdots$& $\vdots$& $\vdots$\\
   \:\:\:\dots\:   \f{1} &   \dcz       &   \dcz       &   \f{1}         &   \dcz       & \dcz\\
  \dots\:\:  \dcz      &  \f{1}  &   \dcz       &   \0  &   \f{1}       & \dcz \\
   \dots\:\: \dcz      &  \dcz        &  \f{1} &   \0        & \0  & \f{1}  \\
   \dots\:\: \dcz      &  \dcz        &  \dcz        &   \0       &   \0     & \0 \\
\end{tabular}
\hspace{8pt}
}
\subfloat{
\tiny
\begin{tabular}[c]{ c c c}
 $\:\:\:\quad\vdots$ & $\vdots$&  $\vdots$\\
    \:\:\:\dots\:   \f{1}  &   \dcz  &   \dcz\\
   \dots\:\:   \dcz  &   \f{1}  &   \dcz\\
  \dots\:\:    \dcz  &   \dcz  &   \f{1}\\
  \dots\:\:    \dcz  &   \dcz  &   \dcz\\
\end{tabular} 
\hspace{-10pt}
}
$\left. \vphantom{\begin{cases} a\vspace{38pt}\end{cases}} \right\rfloor$
\\[8pt]  \small (a) Case 1 \hspace{140pt} (b) Case 2\hspace{185pt} (c)
\vspace{-5pt}
\caption{\small The assigned 1's are boxed, the indeterminate zeros are shown as $\dcz$, and the structural zeros are underlined. (a,b): The two possible cases addressed in Algorithm \ref{alg::one} for allocation of the 1-diagonal in  the $k$-th block-column, separated by the solid vertical line from the previous block-column; 
(a) No structural zero are present along the trajectory of the 1-diagonal extended from the previous block-column; 
(b) Structural zeros force breaking of the 1-diagonal: 
the assigned 1-diagonal starts just above the structural zeros.
Panel (c) shows an example of the undesired case for the last block-column, wherein  the last row is not covered. 
}
\label{fig::AlgCases_SuffProof}
\vspace{-0.2in}
\end{figure}

\vspace{-0.2in}
\subsection{Proof of the sufficient-length proposition} 
\label{sec::SuffCond_proof}

We start by noting that, by Lemma \ref{thm::PRCond}, a sufficient condition for PR is that all $\bH_p$, $p=0,\dots,D-1$, have full row rank.  Hence,  the following result implies   Proposition \ref{thm::FullRankGenericSuff}. 
\vspace{-6pt}
\newtheorem{theorem_StrongFullRankGenericSuff}[theorem]{Proposition}
\begin{theorem_StrongFullRankGenericSuff}\label{thm::StrongFullRankGenericSuff}
For a $D$-fold subsampled $C$-channel length-$m_h$ FIR analysis bank with $C/D\geq 2$, the following property holds generically\textsc{:}  
the sampling-domain analysis polyphase matrix $\bH_p$ corresponding to synthesis filter lengths $m_v\geq m_v^{\st S}(C,D,m_h)$ has full row rank, for all \mbox{$p=0,\dots,D-1$.} 
\end{theorem_StrongFullRankGenericSuff}

Our main tool for proving this proposition is the following result, which provides a test for a matrix function to generically have full rank.

\vspace{-6pt}
\newtheorem{lemma_haribreslerTSP98}{Theorem}
\begin{lemma_haribreslerTSP98}\label{lem::haribreslerTSP98}  
(Harikumar and Bresler \cite{HarBre98a})\\ Let $A(u)$ be an $m\times n$ complex matrix function with elements $A_{i,j}(u)$ that are multivariate polynomials in the elements of $u\in\mbb{C}^{k}$. Then, $A(u)$ has full column rank for almost all $u\in\mbb{C}^{k}$ if it has full column rank for at least one $u\in\mbb{C}^{k}$.
\end{lemma_haribreslerTSP98}
\vspace{-6pt}

The main idea behind this theorem \cite{HarBre98a} is to establish a connection between generic full rank property of structured matrices and \emph{algebraic sets} \cite{Ful69} in the Euclidean space of variables.\footnote{A set is called an algebraic set if it can be written as the set of common zeroes of a system of polynomials \cite{Ful69}.} The result follows by noting that all algebraic sets have zero Lebesgue measure in the Euclidean space.

To proceed with the proof of Proposition \ref{thm::StrongFullRankGenericSuff}, we apply Theorem \ref{lem::haribreslerTSP98} with the variable vector defined as $u=\mathrm{vec}\big\{[h_1, \dots, h_C]\big\}$ and $A(u) = \bH_p^{\mathrm{T}}$ (for each $p=0,\dots,D-1$). It is easy to see that the entries of $\bH_p$ are polynomials of order zero or one in $u$. Therefore, to apply Theorem \ref{lem::haribreslerTSP98}, we need to \emph{construct a particular matrix}, with the same structure as $\bH_p^{\mathrm{T}}$, that has full column rank. In other words, we need to find a set of analysis filters 
$\{h_i\}_{i=1}^{\scsc C}$ for which $\bH_p$ has full row rank.


The following lemma provides the basic idea behind construction of such a matrix.  
The proof is provided in Appendix \ref{sec::app_proof_mvstar}. 

\vspace{-6pt}
\newtheorem{lemma_construction_idea}[lemma]{Lemma}
\begin{lemma_construction_idea}\label{lem::ConstructionIdea}  
Matrix $A$ has full row rank if: (i) each column has at most one nonzero element; (ii) each row has at least one nonzero element. 
\end{lemma_construction_idea}
\vspace{-6pt}

To use Lemma \ref{lem::ConstructionIdea}, we construct an analysis polyphase matrix $\bH_p$ (corresponding to a certain choice of the analysis bank) that   possesses the two properties listed in the lemma. 
%
%
Given the notions of blocks and block-columns described in Section \ref{sec::preliminary_polyph}, let us consider the analysis polyphase matrix constructed by  Algorithm \ref{alg::one} below. The algorithm sequentially assigns a \emph{single} nonzero diagonal or sub-diagonal, called a \emph{1-diagonal}, to \emph{each} block-column; here, a row of $\bH_p$ is called \emph{covered} if at least one of its entries is assigned to be 1. 

\begin{algorithm}
\caption{~Matrix construction for proof of Proposition \ref{thm::StrongFullRankGenericSuff}.}
\label{alg::one}
\medskip
\small
\begin{itemize}
\item[(i)\,--\!] Initialize: $\bH_p = 0$ by assigning 0 to all free entries
\item[(ii)\,--\!]  For the first block-column, assign 1's to the top diagonal (corresponding to  $h_{1,0}[0] =1$) 
\item[(iii)\,--\!]  
While the last row of $\bH_p$ is not covered, assign the 1-diagonal in the $k$-th block-column, $k=2,\dots,C$, according to the following procedure: 
\vspace{-2pt}
\begin{itemize}
\item[$\bullet$] Case 1: If no structural zero is present along the trajectory of the 1-diagonal extended from the previous block-column, assign the 1-diagonal   (for the $k$-th block-column) such that the one in the previous block-column is extended, as shown in Fig.~\ref{fig::AlgCases_SuffProof}(a). 
\item[$\bullet$] Case 2: Otherwise, 
%
assign the 1-diagonal (for the $k$-th block-column) such that it starts \emph{immediately above} the structural zeros in its first column --- two examples of this case are shown in Fig.~\ref{fig::AlgCases_SuffProof}(b). 
\end{itemize}
\item[(iv)\,--\!]  If the last row of $\bH_p$ (last row of $C$-th block-column) is not covered declare \texttt{Failure}.
\item[]
\vspace{-10pt}
\end{itemize}
\end{algorithm}


Figure \ref{fig::Example_Suff_proof} shows the full row rank matrix constructed using Algorithm \ref{alg::one} corresponding to an analysis FB with the following specifications: $D = 3$ subsampling, $C = 6$ channels, and filter length $m_h = 7$. The synthesis filter length $m_v$ is taken to be $m_v^{\st S}(C,D,m_h)$, the minimal value that would satisfy the sufficient-length condition in \eqref{eq::suffcond_condition}, i.e., $m_v = 6$. The assigned 1's are boxed and the structural zeros are underlined. 
The corresponding analysis bank is:
~$h_1[n] = \delta_0(7)$, 
$h_2[n] = \delta_6(7)$, 
$h_3[n] = \delta_1(7)$, 
$h_4[n] = \delta_4(7)$, 
$h_5[n] = \delta_2(7)$, 
$h_6[n] = \delta_5(7)$. 

As suggested by the example given in Fig.~\ref{fig::Example_Hp_Matrix}, there are two factors in the structure of $\bH_p$ that one needs to observe when allocating the nonzero entries: (i) the Toeplitz structure of the blocks; and (ii) structural zeros that limit the number of \emph{free} (assignable) entries. 
The idea in Algorithm \ref{alg::one} is to assign a single 1-diagonal  to each block-column, which corresponds to assigning only one nonzero tap to each analysis filter. It is easy to see that this strategy ensures that the constructed $\bH_p$ would enjoy Property (i) in Lemma \ref{lem::ConstructionIdea} --- hence, all we need to be concerned with is Property (ii). 

The algorithm starts by initializing the matrix to be all zeros and assigning 1's to the top diagonal in the first block-column.  Assume that we have assigned   all of the   entries in the first $k-1$ block-columns, $k=2,\dots, C$. For the $k$-th block-column, which corresponds to $h_k[n]$, there are two possible cases in terms of the structural zeros, as shown in Fig.~\ref{fig::AlgCases_SuffProof}(a,b). Panel (a) shows the simple case, i.e., when no structural zero are present along the trajectory of the 1-diagonal assigned for the $(k-1)$-th block-column. That is, we can simply extend the diagonal trace of 1's by assigning the nonzero entry of the $k$-th column block accordingly. In the example given in Fig.~\ref{fig::Example_Suff_proof}, this case applies to all but the 4\textsuperscript{th} and  6\textsuperscript{th} block-columns, where the structural zeros prevent the extension procedure just described. Two general instances for such nontrivial cases are shown in Fig.~\ref{fig::AlgCases_SuffProof}(b).  To satisfy Property (ii) in Lemma \ref{lem::ConstructionIdea}, we have to assign at least one nonzero entry to each row. As described in Fig.~\ref{fig::AlgCases_SuffProof}(b), this is accomplished by assigning the 1-diagonal for  the $k$-th block-column to start at the last free entry in its first column, i.e., just above the structural zeros. 

According to Algorithm \ref{alg::one}, the rows in a block-row are progressively covered --- i.e., in each step, the non-covered rows are at the bottom of the block-row. Hence, covering of a block-row implies that the last row is covered. 
%
Assume that the assigned nonzero elements used to cover these rows are located in column-blocks $k_1$ to $k_2<C$. 
Then, the filter taps in block-column $(k_2+1)$ are assigned to cover the first $m_{v;p}$ rows of the next block-row without any overlap with those used to cover the previous block-row --- in short,  Algorithm \ref{alg::one} does not ``revisit'' any block-rows.\footnote{Note that the top left element of each Toeplitz block cannot be a structural zero.} 

Consequently, what remains to be shown (to establish Property (ii) in Lemma \ref{lem::ConstructionIdea}) is that there are enough block-columns (out of a total of $C$) to cover all rows of the constructed $\bH_p$ matrix.  
Figure \ref{fig::AlgCases_SuffProof}(c) shows a possible scenario where the last block-column fails to cover all rows (the bracket on the right-hand side shows the bottom-right edge of the matrix). 
We  first show this  for $p=\po$ where, in accordance with Property (c) in \eqref{eq::polyphlength_properties}, $\po$ is such that $m_{v;\po} = \lfloor m_v/D\rfloor.$ 
Let us recall the assumption in the statement of Proposition \ref{thm::StrongFullRankGenericSuff}, $m_v\geq m_v^{\st S}(C,D,m_h)$. 
As pointed out earlier, this means that $m_v$   satisfies \eqref{eq::suffcond_condition}, which can be equivalently written as follows:
\beqn
\sum_{\ell=0}^{D-1} \Bigg\lceil\frac{m_{h;\ell} + m_{v;\po}  - 1}{m_{v;\po}}\Bigg\rceil  \leq C.
\label{eq::suffcond_proof}
\eeqn

Now, consider the following lemma (proved in Appendix \ref{sec::app_proof_mvstar}). 
\vspace{-6pt}
\newtheorem{lemma_construction_rowcovering}[lemma]{Lemma}
\begin{lemma_construction_rowcovering}\label{lem::ConstructionRowCovering}  
Using Algorithm \ref{alg::one}, the number of consecutive block-columns required to cover the \mbox{$m_{h;\ell} + m_{v;p}  - 1$} rows of the $\ell$-th block-row in $\bH_p$ ($0\leq \ell \leq D-1$) is equal to $\big\lceil(m_{h;\ell} + m_{v;p}  - 1)/m_{v;p}\big\rceil$. 
\end{lemma_construction_rowcovering}
\vspace{-3pt}
\noindent By Lemma \ref{lem::ConstructionRowCovering}, Inequality \eqref{eq::suffcond_proof} guarantees that the $C$ block-columns  of $\bH_\po$ suffice for the algorithm to cover all rows of $\bH_{\po}$ --- i.e., avoid the case shown in Fig.~\ref{fig::AlgCases_SuffProof}(c). 
Finally, since by \eqref{eq::polyphlength_properties} we have $m_{v;p}\geq m_{v;\po}$, \eqref{eq::suffcond_proof} implies that the same argument holds for all $\bH_p$ with $0\leq p \leq D-1$. 
As a result, the condition in Line (iv) of Algorithm \ref{alg::one} is not met for any $p$; hence, the algorithm successfully finishes the construction of $\bH_p$, $0\leq p \leq D-1.$ 	  
\hfill $\square$

\section{Minimum Length for PR Synthesis Filters: Generic Necessary Condition}\label{sec::NecCond}

In the previous section, we provided a ``sufficient length'' condition for PR synthesis banks. However, this only partially answers Question Q.2 (Section \ref{sec::intro}) as it raises the possibility that the prescribed minimum length $m_v^{\st S}(C,D,m_h)$ is too conservative. 
To refute this possibility, in this section, we propose a \emph{necessary} condition counter-part to Proposition \ref{thm::FullRankGenericSuff}. It states that for synthesis filter lengths below a certain necessary length $m_v^{\st N}$ (defined below) PR or delayed PR cannot  generically be achieved. 
Subsequently, we   exactly quantify the gap between the sufficient and the necessary lengths for each choice of $(C,D,m_h)$.  
In Section \ref{sec::bounds_n_relations}, we provide numerical results demonstrating that the gap between these two lengths is indeed small (for moderately high oversampling factors).  



\vspace{-10pt}
\subsection{Statement of the result} \label{sec::NecCondstatement}

First, let us define the counterpart to the sufficient length $m_v^{\st S}(C,D,m_h)$. 

\vspace{-8pt}
\begin{definition}
Denote by $m^{\st N}_v(C,D,m_h)$ the minimal value of $m_v\in\mathbb{N}$ that satisfies:
\beqn
\frac{C}{D} \geq 1 + \frac{1}{D}\sum_{p=0}^{D-1} \Bigg\lfloor\frac{m_{h;p}  - 1}{\lfloor m_v/D \rfloor}\Bigg\rfloor, 
\label{eq::neccond_condition}
\eeqn
where $m_{h;p} = \Big\lceil \frac{m_h - p}{D} \Big\rceil$. We refer to $m^{\st N}_v$ as the necessary synthesis filter length --- in short, the \emph{necessary length}.
\end{definition}
\vspace{-6pt}
It is easy to show that all $m_v\geq m^{\st N}_v(C,D,m_h)$ satisfy \eqref{eq::neccond_condition}. Therefore, the set of $m_v\in\mbb{N}$ satisfying \eqref{eq::neccond_condition} is a right-sided interval, i.e., all integers in $\Big[m_v^{\st N}(C,D,m_h), \infty\Big)$. 
Furthermore,   the only difference between the definitions of the necessary   and sufficient lengths (Section \ref{sec::SuffCond}) is a floor/ceiling operation in the summand. 

The following proposition, the main result of this section, provides a necessary condition counterpart to Proposition \ref{thm::FullRankGenericSuff} for PR in generic FBs. 

\vspace{-5pt}
\newtheorem{theorem_nec}[theorem]{Proposition}
\begin{theorem_nec}\label{thm::FullRankGenericNec}  
\textbf{\textsc{(}necessary length\textsc{)}}~
A $D$-fold subsampled $C$-channel length-$m_h$ FIR analysis bank with $C/D\geq2$ and $m_h>D$ is generically not invertible, i.e., the FB does not admit PR with any delay $n_0$,  $0\leq \lceil\frac{n_0}{D}\rceil \leq \lfloor \frac{m_h}{D} \rfloor+\lfloor \frac{m_v}{D} \rfloor-2$,  by a synthesis FB of length $m_v < m_v^{\st N}(C,D,m_h)$. 
\end{theorem_nec}
\vspace{-10pt}

\vspace{-3pt}
\subsection{Proof of of the necessary-length proposition}
 \label{sec::NecCondproof}

%


The condition $m_v<m_v^{\st N}(C,D,m_h)$  implies that  \eqref{eq::neccond_condition} is not satisfied (since the set satisfying \eqref{eq::neccond_condition}  is a right-sided interval as pointed out earlier). Here, we show that violating \eqref{eq::neccond_condition} in turn implies that generically   there exists an integer $p\in\{0,\dots,D-1\}$ such that $\vv \delta_{\kappa(p,n_0)} \notin \Range (\bH_p)$ for any $n_0$, which by Lemma \ref{thm::PRCond}  is equivalent to the proposition.   
To this end, we construct the augmented matrix $\At_p = \big[\bH_p \:\:\:\: \vv\delta_{\kappa(p,n_0)} \big]$, and will establish that $\At_p$ is generically \emph{full column rank}, for some $p$. 

To proceed with the proof, we apply Theorem \ref{lem::haribreslerTSP98} with $A(x) = \At_p$ and $ x=\mathrm{vec}\big\{[h_1, \dots, h_C]\big\}$. The entries of $\At_p$ are polynomials of order zero or one in $x$. Hence, if we find a particular set of analysis filters 
$\{h_i\}_{i=1}^{\scsc C}$ such that the corresponding $\At_p$ matrix has full column rank, then it will be generically full column rank, which in turn proves the proposition. 
Accomplishing this is equivalent to constructing a sampling-domain analysis polyphase matrix $\bH_p$ (for at least one $p$) such that: (a) $\bH_p$ has full column rank; (b) $\vv\delta_{\kappa(p,n_0)}$ is linearly independent of all columns of $\bH_p$. In what follows, the matrix construction corresponds to $p=\po$, where $\po$ was defined in Property (c) of \eqref{eq::polyphlength_properties}.


Let us consider the matrix $\bH_\po$ constructed by Algorithm \ref{alg::two} below.  
%
%
%
Note that for the assignment in Line (ii) of the algorithm to be feasible we need $m_{h;0}>m_{v;\po}$, which is guaranteed by the following lemma.\footnote{For  $h_{1,0}[m_{v;\po}] = 1$ to be feasible the length of $h_{1,0}$ should be at least $m_{v;\po}$+1.}   
The proof is provided in Appendix \ref{sec::app_proof_mvstar}. 
\vspace{-8pt}
\newtheorem{lemma_Line2_Alg2}[lemma]{Lemma}
\begin{lemma_Line2_Alg2}\label{lem::Line2_Alg2}  
With the assumptions in Proposition \ref{thm::FullRankGenericNec}, we have $m_{h;0}>m_{v;\po}$.
\end{lemma_Line2_Alg2}
\vspace{-4pt}
%
\begin{algorithm}
\caption{~Matrix construction for proof of Proposition \ref{thm::FullRankGenericNec}.}
\label{alg::two}
%
%
%
\medskip
\small
\begin{itemize}
\item[(i)\,--\!]  Initialize: $\bH_\po = 0$ by assigning 0 to all free entries.
\item[(ii)\,--\!]  For the first block-column, select the 1-diagonal to be the $(m_{v;\po}+1)$-th diagonal in the first Toeplitz block (corresponding to $h_{1,0}[m_{v;\po}] = 1$).
\item[(iii)\,--\!]  For the $k$-th block-column, $k=2,\dots,C$, select the 1-diagonal according to the following procedure: 
\vspace{-4pt}
\begin{itemize}
\item[$\bullet$] Case 1: If no structural zero is present along the trajectory of the 1-diagonal extended from the previous block-column, assign the 1-diagonal   (for the $k$-th block-column) such that the one in the previous block-column is extended, as shown in Fig.~\ref{fig::AlgCases_NecProof}(a).
\item[$\bullet$] Otherwise,
\vspace{-4pt}
\begin{itemize}
\item[---]  Case 2: Assign the 1-diagonal (for the $k$-th block-column) such that it starts \emph{immediately below} the structural zeros in its first column --- an example of this case is shown in Fig.~\ref{fig::AlgCases_NecProof}(b). 
\item[---]  Declare \texttt{Failure} if there are no free entries left below the structural zeros, e.g., as shown in Fig.~\ref{fig::AlgCases_NecProof}(c).
\end{itemize}
\end{itemize}
\item[(iv)\,--\!]  Let $c^{\ast}$ be the block-column index of the nonzero entry in the row indexed by $\kappa(\po,n_0)$.
\item[(v)\,--\!]  If all entries of the row indexed by $\kappa(\po,n_0)$ are zero then \texttt{Exit};\\ Else, 
in the $c^{\ast}$-th block-column, assign the top diagonal to be nonzero (corresponding to $h_{c^{\ast},0}[0] = 1$). 
\smallskip
\end{itemize}
\end{algorithm}
%
%
Figure \ref{fig::Example_Nec_proof} provides an example of the matrix $\bH_\po$ with $\po=2$ constructed by Algorithm \ref{alg::two} for a FB with $C=6$ channels, $D=3$ subsampling, and analysis filter length of $m_h = 13$. The reconstruction delay is $n_0=6$ (corresponding to $\kappa(\po,n_0)=13$).  In accordance with the assumption $m_v<m_v^{\st N}(C,D,m_h)$, we have $m_v =  m_v^{\st N}(C,D,m_h) - 1 = 8$. 
 The arrow next to $\bH_2$ indicates the row indexed by $\kappa(\po,n_0)$, which is the location of the 1 in $\vv\delta_{\kappa(\po,n_0)}$.  
The corresponding analysis bank is:
$h_1[n] = \delta_6(13)$, 
$h_2[n] = \delta_{12}(13)$, 
$h_3[n] = \delta_1(13)$, 
$h_4[n] = \delta_7(13)$, 
$h_5[n] = \delta_2(13)$, 
$h_6[n] = \delta_0(13) +  \delta_8(13)$. 

\begin{figure}[t!]
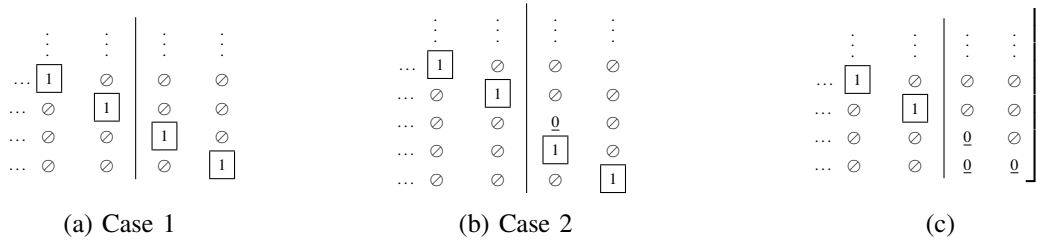

\centering
\subfloat{
\hspace{-6pt}
\tiny
\begin{tabular}[c]{ c c | c c }
    $\:\:\:\quad\vdots$ & $\vdots$&  $\vdots$& $\vdots$\\
    \:\:\:\dots\:\f{1}  &   \dcz  &   \dcz  &   \dcz  \\
    \dots\:\:  \dcz  &   \f{1}  &   \dcz  &   \dcz  \\
   \dots\:\:   \dcz  &   \dcz  &   \f{1}   & \dcz   \\
   \dots\:\:   \dcz  &   \dcz  &   \dcz  &   \f{1}  \\
\end{tabular}
\hspace{38pt}
}
\subfloat{
\tiny
\begin{tabular}[c]{ c c | c c }
    $\:\:\:\quad\vdots$ & $\vdots$&  $\vdots$& $\vdots$\\
    \:\:\:\dots\: \f{1}  &   \dcz  &   \dcz  &   \dcz \\
     \dots\:\: \dcz  &   \f{1}  &   \dcz  &   \dcz  \\
     \dots\:\: \dcz  &   \dcz  &        \0  &   \dcz \\
        \dots\:\: \dcz  &   \dcz  &     \f{1}  &   \dcz  \\
                \dots\:\: \dcz  &   \dcz  &   \dcz &  \f{1}   \\
\end{tabular} 
\hspace{48pt}
}
\subfloat{
\tiny
\begin{tabular}[c]{ c c | c c }
    $\:\:\:\quad\vdots$ & $\vdots$&  $\vdots$& $\vdots$\\
    \:\:\:\dots\: \f{1}  &   \dcz  &   \dcz  &   \dcz \\
     \dots\:\: \dcz  &   \f{1}  &   \dcz  &   \dcz  \\
     \dots\:\: \dcz  &   \dcz  &     \0  &   \dcz  \\ 
      \dots\:\: \dcz  &   \dcz  &   \0&  \0   \\
\end{tabular} 
\hspace{-10pt}
}
$\left. \vphantom{\begin{cases} a\vspace{42pt}\end{cases}} \right\rfloor$
\\[8pt]  \small (a) Case 1 \hspace{100pt} (b) Case 2\hspace{130pt} (c)
\vspace{-5pt}
\caption{\small 
(a,b): The two possible cases addressed in Algorithm \ref{alg::two} for allocation of the 1-diagonal for  the $k$-th block-column, separated by the solid vertical line from the $(k-1)$-th block-column; (a) No structural zero are present along the trajectory of the 1-diagonal
; (b) Structural zeros force breaking of the 1-diagonal: 
the 1-diagonal for the $k$-th block-column is chosen such that it would start just below the structural zeros in its first column. Panel (c) shows an example of the undesired case for the last block-column, wherein there are no free entries left below the structural zeros.   
}
\label{fig::AlgCases_NecProof}
\vspace{-0.25in}
\end{figure}
\begin{figure}[t!]
\centering
\begin{minipage}[b]{1\linewidth}
\beqn
\bH_2 =
\tiny
\left[
\begin{array}{c c | c c | c c  | c c | c c | c c }
0  &   \0  &   0  &  \0  &   0  &  \0  &  0  &   \0  &  0  &  \0  &   \f{1}  &  \0 \\
0  &   0  &   0  &  0  &   0  &  0  &  0  &   0  &   0  &  0  &   0  &  \f{1} \\
\f{1}  &   0  &   0  &  0  &   0  &  0  &  0  &   0  &   0  &  0  &   0  &  0 \\
0  &   \f{1}  &   0  &  0  &   0  &  0  &  0  &   0  &   0  &  0  &   0  &  0 \\
0  &   0  &   \f{1}  &  0  &   0  &  0  &  0  &   0  &   0  &  0  &   0  &  0 \\
\0  &   0  &   \0  &  \f{1}  &   \0  &  0  &  \0  &   0  &  \0  &  0  &   \0  &  0 \\[1pt]
\hdashline
0  &   \0  &   0  &  \0  &   \f{1}  &  \0  &  0  &   \0  &   0  &  \0  &   0  &  \0 \\
0  &   0  &   0  &  0  &   0  &  \f{1}  &  0  &   0  &   0  &  0  &   0  &  0 \\
0  &   0  &   0  &  0  &   0  &  0  &  \f{1}  &   0  &   0  &  0  &   0  &  0 \\
0  &   0  &   0  &  0  &   0  &  0  &  0  &   \f{1}  &   0  &  0  &   0  &  0 \\
\0  &   0  &   \0  &  0  &   \0  &  0  &  \0  &   0  &   \0  &  0  &   \0  &  0 \\[1pt]
\hdashline
0  &   \0  &   0  &  \0  &   0  &  \0  &  0  &   \0  &    \f{1}  &  \0  &   0  &  \0 \\
0  &   0  &   0  &  0  &   0  &  0  &  0  &   0  &   0  &  \f{1}  &   0  &  0 \\
0  &   0  &   0  &  0  &   0  &  0  &  0  &   0  &   0  &  0  &   \f{1}  &  0 \\
0  &   0  &   0  &  0  &   0  &  0  &  0  &   0  &   0  &  0  &   0  &  \f{1} \\
\0  &   0  &   \0  &  0  &   \0  &  0  &  \0  &   0  &  \0  &  0  &   \0  &  0 \\
\end{array}
\right]
\begin{array}{ll}
   {}  \\
   {}  \\
   {}  \\
   {}  \\
   {}  \\
   {}  \\
   {}   \\
   {}  \\
   {}  \\
   {}  \\
   {}  \\
   {} \\
   {}  \\[5pt]
   \myArrow  \\
   {}  \\
   {}  \\   
 \end{array} 
\nonumber 
\eeqn
\end{minipage}
\caption{\small The analysis polyphase matrix $\bH_\po$ (size: $16\times 12$) for $\po=2$ and $n_0=6$ delay (corresponding to $\kappa(\po,n_0)=13$) constructed by Algorithm \ref{alg::two} for a 6-channel FB with $D=3$ and $m_h = 13$ (hence, $m_{h;0}=5, m_{h;1}=m_{h;2}=4$). Based on the assumption in the proof, the synthesis filter length is taken to be $m_v = m_v^{\st N}(C,D,m_h) - 1 = 8$ (hence, $m_{v;2}=2$). The arrow next to $\bH_2$ indicates the location of the 1 in $\vv\delta_{\kappa(\po,n_0)}$.}
\label{fig::Example_Nec_proof}
\vspace{-0.25in}
\end{figure}


The first task is to prove that the constructed $\bH_\po$ has full column rank. This is accomplished by showing that it has the two properties given in the following lemma, which is equivalent to Lemma \ref{lem::ConstructionIdea}. 
\vspace{-6pt}
\newtheorem{lemma_construction_idea2}[lemma]{Lemma}
\begin{lemma_construction_idea2}\label{lem::ConstructionIdea2}  
Matrix $A$ has full column rank if\textsc{:} (i) each row has at most one nonzero element; (ii) each column has at least one nonzero element.
\end{lemma_construction_idea2}
\vspace{-8pt}


%


By construction, Algorithm \ref{alg::two}: (a) allocates at most a single 1-diagonal to each Toeplitz block in $\bH_\po$;
(b) if the 1-diagonal for some block-column has a maximum row index of $r_\circ$, then the next block-column would not have any nonzero entries at or above the $r_\circ$-th row. 
These two observations together imply Property (i) above. 

Next, we prove that Property (ii) in Lemma \ref{lem::ConstructionIdea2}  holds. 
We call a block-column \emph{covered} (by Algorithm \ref{alg::two}) if all of its  $m_{v;\po}$ columns satisfy Property (ii).  
Owing to the Toeplitz structure of the blocks,  any 1-diagonal assignment by the algorithm corresponds to covering of a block-column. However, it could be that, at a certain iteration of the loop in Line (iii) of the algorithm, there would be no free entries (that are below the structural zeros) left to assign. Figure \ref{fig::AlgCases_NecProof}(c) shows an example of such a scenario for the last ($C$-th) column-block. In the following we show that, given the  assumption of the proof, i.e., $m_v<m_v^{\st N}(C,D,m_h)$, this latter scenario will never materialize.  
Consider the following lemma. (The proof is analogous to that of Lemma \ref{lem::ConstructionRowCovering} and is therefore skipped.)
\vspace{-5pt}
\newtheorem{lemma_construction_rowcovering2}[lemma]{Lemma}
\begin{lemma_construction_rowcovering2}\label{lem::ConstructionRowCovering2}  
\quad Using Algorithm \ref{alg::two}, the number of consecutive block-columns in $\bH_\po$ covered by   the \mbox{$m_{h;\ell} + m_{v;\po}  - 1$} rows of the $\ell$-th block-row  ($0\leq \ell \leq D-1$) is equal to $\big\lfloor(m_{h;\ell} + m_{v;\po}  - 1)/m_{v;\po}\big\rfloor$. 
\end{lemma_construction_rowcovering2}
\vspace{-4pt}

%
The opposite of \eqref{eq::neccond_condition}, which is implied by the assumption $m_v<m_v^{\st N}(C,D,m_h)$, can be rewritten as:
\beqn
\frac{C}{D} < 1  + \frac{1}{D}\sum_{\ell=0}^{D-1} \Bigg\lfloor\frac{m_{h;\ell}  - 1}{\lfloor m_v/D \rfloor}\Bigg\rfloor  \Rightarrow 
C  - D < \sum_{\ell=0}^{D-1} \Bigg\lfloor\frac{m_{h;\ell}  - 1}{\lfloor m_v/D \rfloor}\Bigg\rfloor
 \Rightarrow 
D + \sum_{\ell=0}^{D-1} \Bigg\lfloor\frac{m_{h;\ell}  - 1}{m_{v;\po}}\Bigg\rfloor \geq C+1,
\label{eq::neccond_negated}
\eeqn
where, in the last step, we assumed $m_{v;\po} = \lfloor m_v/D \rfloor$. 
%
Now, summing up the number of consecutive block-columns that can be covered by all rows, we have:
\beqn
\bigg\lfloor \frac{m_{h;0} - 1}{m_{v;\po}}\bigg\rfloor + \sum_{\ell=1}^{D-1}\bigg\lfloor \frac{m_{h;\ell} + m_{v;\po} - 1}{m_{v;\po}}\bigg\rfloor  = - 1 + \bigg(D + \sum_{\ell=0}^{D-1}\bigg\lfloor \frac{m_{h;\ell} - 1}{m_{v;\po}}\bigg\rfloor \bigg) \geq -1 + C + 1 = C, 
\label{eq::neccond_negated_proof}
\eeqn
where we applied \eqref{eq::neccond_negated}. The inequality in \eqref{eq::neccond_negated_proof} implies that there are enough rows to cover   all $C$ block-columns.  
%
Using a similar argument as the one at the end of Section \ref{sec::SuffCond_proof}, it is easy to see that Algorithm \ref{alg::two} covers the block-columns progressively, i.e., once a block-column is covered it is not revisited. This in combination with \eqref{eq::neccond_negated_proof}  guarantees that the \texttt{Failure} condition in Line (iii) of the algorithm is never met; hence,  the matrix construction finishes successfully --- satisfying both properties in  Lemma \ref{lem::ConstructionIdea2}. 


Finally, we show that  $\vv\delta_{\kappa(\po,n_0)}$ is linearly independent of all columns of the constructed $\bH_\po,$ which --- together with the above --- establishes that the the constructed (augmented) matrix $\At_\po = \big[\bH_\po \:\:\:\: \vv\delta_{\kappa(\po,n_0)} \big]$ is full column rank. 
By inspection, it is clear that the top $m_{v;\po}$ rows of the matrix constructed in Lines (i)--(iv) of Algorithm \ref{alg::two} are set to zero. Consider two cases based on whether $m_{v;\po}$ is larger  or smaller than $\kappa(\po,n_0)$.  
If $\kappa(\po,n_0)<m_{v;\po}$, the condition in Line (v) of Algorithm \ref{alg::two} holds and no linear combination of columns of $\bH_\po$ can produce $\vv\delta_{\kappa(\po,n_0)}$. 
%
For the alternative case of $\kappa(\po,n_0)\geq m_{v;\po}$, the $1$ in $\vv\delta_{\kappa(\po,n_0)}$ may share the same row index with at most one nonzero element in $\bH_\po$ --- denote the corresponding column in $\bH_\po$ by $e$.\footnote{The case where no such element exists is trivial.} 
Line (v) of Algorithm \ref{alg::two} (after ``Else, ...'') ensures that $e$ will have another nonzero element  in the top $m_{v;\po}$ rows of the matrix, i.e., in the first row-block of $\bH_\po$. This is demonstrated in Fig.~\ref{fig::Example_Nec_proof} where $e$ is the eleventh column and $m_{v;\po} =2$.   
Consequently, since $\kappa(\po,n_0)\geq m_{v;\po}$, column $e$ cannot belong to any linear combination producing $\vv\delta_{\kappa(\po,n_0)}$, which in turn implies that $\vv\delta_{\kappa(\po,n_0)}$ is linearly independent of all columns of $\bH_\po$.  
This completes the proof. 
\hfill $\square$

\section{Tightness, Gaps, and Closed-form Expressions} 
\label{sec::bounds_n_relations}

In the previous two sections, we provided generic necessary and sufficient requirements for the length of the PR synthesis bank. 
 Here we  show that the gap between the necessary length $m_v^{\st N}(C,D,m_h)$ and the sufficient length $m_v^{\st S}(C,D,m_h)$ is small.  
%
In addition, to answer the second part of Question Q.2 of Section \ref{sec::intro}, we study the fundamental relation between the oversampling ratio, the analysis filter length, and the required synthesis length. 
This is facilitated by providing various bounds including closed-form upper and lower bounds for the sufficient and necessary synthesis filter lengths, respectively.
%
The derived relations are verified and illustrated numerically  in Section \ref{sec::results_filterlength}.


%
 Assuming that the FB (Fig.~\ref{fig::multichannel}) is oversampled, we start with a simple  necessary condition, referred to as the \emph{counting condition}, for all $\bH_p$, $p=0, \dots, D-1$, to be full row rank.
The proof is provided in Appendix \ref{sec::app_proof_mvstar}.
\vspace{-0.1in}
\newtheorem{CountingCond}[lemma]{Lemma}
\begin{CountingCond}\label{thm::CountingCond}  
\textbf{\textsc{(}counting length\textsc{)}}~
Assuming $C>D$, a necessary condition for all  $\bH_p$ ($p=0,\dots,D-1$) to be full row rank is
\beqn
m_v\geq m^{\st C}_v(C,D,m_h) \defeq D\left\lceil \frac{\obliquefrac{m_h}{D} - 1}{\obliquefrac{C}{D} - 1} \right\rceil,
\label{eq::lemma_counting}
\eeqn
where the integer functional $m^{\st C}_v(C,D,m_h)$ is referred to as the counting length. 
\end{CountingCond}
%
The counting length is simply derived by requiring the number of rows to be no more than the number of columns for all $\bH_p$.    
Note that, because the condition in Lemma \ref{thm::PRCond} may be satisfied even when none of $\{\bH_p\}_{\scsc p=0}^{\scsc D-1}$ have full rank,  condition \eqref{eq::lemma_counting} is neither necessary nor sufficient for existence of a length-$m_v$ PR synthesis bank.  
Remarkably, as described below (also revisited in Section \ref{sec::results_filterlength}), $m^{\st C}_v(C,D,m_h)$  is closely related to $m^{\st S}_v(C,D,m_h)$ and to $m^{\st N}_v(C,D,m_h)$, derived earlier.  

Applying the basic inequality $\alpha \leq \lceil \alpha\rceil < \alpha+1$, we have 
\beqn
\frac{m_h - D}{C/D - 1}\leq m_v^{\st C}(C,D,m_h) < D + \frac{m_h-D}{C/D-1},
\label{eq::lemma_countingBounds}
\eeqn
i.e.,   $m_v^{\st C}(C,D,m_h)$ is bounded from below and above by functions that each have an approximate inverse relationship to $C/D$, for a fixed $m_h$. %
%
%
Since the gap between the bounds in \eqref{eq::lemma_countingBounds} is small (equal to $D$), the counting length $m_v^{\st C}(C,D,m_h)$ has an approximate inverse relationship to the oversampling factor $C/D$. 

The following proposition describes 
%
the relationship between the necessary, sufficient, and counting filter lengths; it further provides closed-form lower/upper bounds for  $m_v^{\st S}$ and $m_v^{\st N}$.
The proof is provided in Appendix \ref{sec::app_proof_mvstar}.
%
%

%
%
%
\vspace{-0.05in}
\newtheorem{theorem_alltogether}[theorem]{Proposition}
\begin{theorem_alltogether}\label{lem::corollary_one} 
Assuming $C/D \geq 2$ and $m_h> D$, we have the following relations between the sufficient length $m^{\st S}_v(C,D,m_h)$, the necessary length $m^{\st N}_v(C,D,m_h)$, and the counting length $m^{\st C}_v(C,D,m_h)$:
$$\max\big(D,m_v^{\st L}\big)\leq m_v^{\st N} \leq m_v^{\st C} \leq m_v^{\st S} \leq \min(m_h, m_v^{\st U}),$$
where $~\displaystyle m_v^{\st L}(C,D,m_h) \defeq \bigg\lceil\frac{m_h-D}{C/D}\bigg\rceil~$ and $~\displaystyle m_v^{\st U}(C,D,m_h) \defeq D + \bigg\lceil\frac{m_h-D}{(C+1)/D-2}\bigg\rceil$. 
\end{theorem_alltogether}
\smallskip

In order to draw conclusions from Proposition  \ref{lem::corollary_one}  in terms of the behavior of the various length, we first need to quantify the gaps between them --- as given in the following corollary. 
The proof is provided in Appendix \ref{sec::app_proof_mvstar}.
%
%
%
\newtheorem{corollary_GapBounds}{Corollary}
\begin{corollary_GapBounds}\label{lem::GapBounds}
Assuming $D\geq 2$ and $C/D\geq 2$, the following bounds apply for the gaps between various length functionals: 
\vspace{-0.05in}
\beqn
\Gamma_{\st S|\st N}(C,D,m_h) &\!\!\!\!\defeq\!\!\!\!& m_v^{\st S}(C,D,m_h) - m_v^{\st N}(C,D,m_h)  <  D+1 + \frac{2 m_h}{\frac{C}{D}\Big( \frac{C}{D-0.5} - 2\Big)} 
\nonumber\\
\Gamma_{\st U|\st C}(C,D,m_h) &\!\!\!\!\defeq\!\!\!\!& m_v^{\st U}(C,D,m_h) - m_v^{\st C}(C,D,m_h)  < D+1 + \frac{m_h}{\big(\frac{C}{D} - 1\big) \big(\frac{C-1}{D-1} - 2\big)}
\nonumber\\
\Gamma_{\st C|\st L}(C,D,m_h) &\!\!\!\!\defeq\!\!\!\!& m_v^{\st C}(C,D,m_h) - m_v^{\st L}(C,D,m_h)  < D+1 + \frac{m_h}{\frac{C}{D}\big( \frac{C}{D} - 1\big)} .
\nonumber
\eeqn
\end{corollary_GapBounds}
\smallskip

For a generic $C$-channel $D$-fold subsampled FB, the \emph{true minimal} PR synthesis filter length, i.e., where the ``phase transition'' between PR-infeasibility and PR-feasibility occurs, is denoted by $m_v^{\ast}(C,D,m_h)$. It follows that  $m_v^{\st N} \leq m_v^{\ast} \leq m_v^{\st S}$.  
%
Although our results do not exactly pinpoint the true minimal filter length $m_v^{\ast}$  (see Section \ref{sec::results_filterlength} for a conjecture that it coincides with $m_v^{\st C}$), we can exactly quantify the gap between $m_v^{\st N}$ and $m_v^{\st S}$ using their respective definitions.  
Here instead, combining Corollary \ref{lem::GapBounds} and Proposition \ref{lem::corollary_one}, we  study  the \emph{qualitative behavior} of the necessary and sufficient lengths 
and the gap between the two, which in turn enables us to address Question Q.2 raised in Section \ref{sec::intro} regarding $m_v^{\ast}$. \\[4pt]
$\bullet$ \emph{The gap between $m_v^{\st N}$ and $m_v^{\st S}$ is small}. 
This is because, based on Corollary \ref{lem::GapBounds}, $\Gamma_{\st S|\st N}(C,D,m_h)$ drops roughly as $(C/D)^{-2}$ and is small (relative to $m_h$) for moderately high oversampling factors. For example, with $D=3$, for $C/D\geq 3$, the gap is smaller than $4 + 0.42\times m_h$, and for $C/D\geq 4$ it is smaller than $4+0.18\times m_h$. 
Further, in Section \ref{sec::results_filterlength}, we illustrate  numerically that this gap is small.  
%
\\[4pt] $\bullet$ \emph{$m_v^{\st S}$ has an approximately inverse relation to the oversampling factor.} 
Consider the set of inequalities in Proposition \ref{lem::corollary_one} bounding the sufficient length: \mbox{$m_v^{\st C} \leq m_v^{\st S} \leq m_v^{\st U}$}. For a fixed analysis filter length $m_h$=$m_h^0$ and subsampling factor $D$=$D_0$, this relation shows that the integer function $m_v^{\st S}(C,D_0,m_h^0)$, defined on the integer line $C\in\mbb{N}$, is ``sandwiched'' between two other integer functions both of which roughly drop as $1/C$ with increasing $C$. 
Moreover, the gap between $m_v^{\st U}$ and $m_v^{\st C}$, denoted as $\Gamma_{\st U|\st C}$ in Corollary \ref{lem::GapBounds}, is small since: (i) it drops as $(C/D)^{-2}$ with increasing $C$; (ii) 
based on the proof of  Proposition \ref{lem::corollary_one} (specifically, Part (i) of Lemma  \ref{lem::FullRankGenericSuffFolk} in Appendix \ref{sec::app_proof_mvstar}), 
this gap is zero for all  $C=kD_0$ with $k\geq 2$.  Therefore, $m_v^{\st S}$ itself should behave similarly. The same argument can be repeated for a fixed $C$ and decreasing $D$. This shows that  $m_v^{\st S}$ has an approximately inverse relation to $C/D$. 
\\[4pt] $\bullet$ \emph{$m_v^{\st N}$ has an approximately inverse relation to the oversampling factor.} Similarly to the above, based on Proposition \ref{lem::corollary_one}, the necessary length  $m_v^{\st N}$ is sandwiched between $m_v^{\st L}$ and $m_v^{\st C}$. Moreover, the gap between $m_v^{\st L}$ and $m_v^{\st C}$, denoted $\Gamma_{\st C|\st L}(C,D,m_h)$ in Corollary \ref{lem::GapBounds}, drops roughly as $(C/D)^{-2}$ and is small. For example, for $C/D\geq 3$, the gap is smaller than $D+1+(m_h/6)$. Specifically, for $m_h=17$ and $D=2$, $\Gamma_{\st C|\st L}\leq 5$. Hence, $m_v^{\st N}(C,D,m_h)$ itself has an approximate inverse relation to $C/D$.  
\\[4pt] $\bullet$ \emph{$m_v^{\ast}$ has an approximate inverse relation to the oversampling factor.}  Summarizing the abovementioned relations, the integer functionals $m_v^{\st N}$ and $m_v^{\st S}$ both behave (approximately) as $(C/D)^{-1}$. Also, the gap in between them drops roughly as $(C/D)^{-2}$ and is small (relative to $m_h$). This implies that  $m_v^{\ast}$, which lies in between $m_v^{\st N}$ and $m_v^{\st S}$, should itself have an approximate inverse relation with respect to the oversampling factor $C/D.$ 
This answers the second part of Question Q.2 in Section \ref{sec::intro} and is further demonstrated in Section \ref{sec::results_filterlength}. 
\smallskip 

%
%
%

The inequality $m_v^{\st S}\leq m_h$ stated in Proposition \ref{lem::corollary_one} implies that, with $C/D\geq 2$, the true minimal PR synthesis filter length $m_v^{\ast}(C,D,m_h)$ is generically less than the analysis filter length $m_h$. 
Note the lower bound $D$ for all lengths, which implies $m_v^{\ast}(C,D,m_h)\geq D$; this is can be inferred from the FB structure (Panel (a) in Fig.~\ref{fig::multichannel}), as a shorter synthesis filter would not be able to ``fill in'' the $D$-sample gap at the output of the up-samplers. 



Finally, let us consider Question Q.1 raised in Section \ref{sec::intro}. 
%
Based on Proposition \ref{thm::FullRankGenericSuff}  (with $m_0=0$), a length-$m_v$ PR synthesis bank exists generically for any $m_v\geq m_v^{\st S}(C,D,m_h)$, where, based on Proposition \ref{lem::corollary_one}, $m_v^{\st S}(C,D,m_h)\leq m_h<\infty$. This implies feasibility of PR using FIR synthesis filters. The following corollary states this observation.
\vspace{-6pt}
\newtheorem{corollary_two}[corollary_GapBounds]{Corollary}
\begin{corollary_two}\label{lem::corollary_two}  
For a $D$-fold subsampled $C$-channel FIR analysis bank with $C/D\geq 2$ and $m_h>D$, the following property holds generically\textsc{:} an FIR 
synthesis bank achieving PR with any $n_0$ delay $\mathrm{\big(}0\leq \lceil\frac{n_0}{D}\rceil \leq \lfloor \frac{m_h}{D} \rfloor+\lfloor \frac{m_v}{D} \rfloor-2\mathrm{\big)}$ exists.
\end{corollary_two}
\vspace{-3pt}


A recent result due to Law et al.~\cite{LawFosDo0911}, specialized to a single-variate polynomial matrix, states that 
if $C\geq D+1$, then a $C\times D$ single-variate polynomial matrix is generically (Laurent) polynomial left invertible.
%
Applying this result to the analysis polyphase matrix $\sH(z)$, we can deduce that for generic oversampled FBs $\sH(z)$ has an FIR left inverse, which corresponds to the synthesis polyphase matrix $\sV(z)$ according to \eqref{eq::polyphPRcond_matrixversion}. Therefore, the result in \cite{LawFosDo0911} implies Corollary \ref{lem::corollary_two} and hence is stronger.
%
%
However, as mentioned in Section \ref{sec::intro}, Law et al.~\cite{LawFosDo0911} do not address the filter support/length question (Q.2 in Section \ref{sec::intro}), which is the focus of the present work.

\section{Numerical Results: Verification of the Propositions and Further Observations
}\label{sec::results_filterlength}

The first part of this section provides numerical verification of the results in Section \ref{sec::bounds_n_relations}, as summarized in Proposition \ref{lem::corollary_one}.
In the second part, we provide Monte-Carlo simulations results that provide a numerical verification of the Propositions \ref{thm::FullRankGenericSuff} and \ref{thm::FullRankGenericNec} in Sections \ref{sec::SuffCond} and \ref{sec::NecCond}, respectively.
A byproduct of the presented numerical results is a conjecture on the true minimal filter length for PR. In the last part of this section, we study the feasibility of imperfect (near perfect) reconstruction using synthesis filter lengths below those prescribed (for PR) by our propositions. 

\begin{figure}[tbp]
\begin{minipage}[b]{1\linewidth}
\centering
\vspace{-0.05in}
\subfloat[]{\includegraphics[trim=35mm 0mm 25mm 0mm, clip, width=0.5\textwidth]{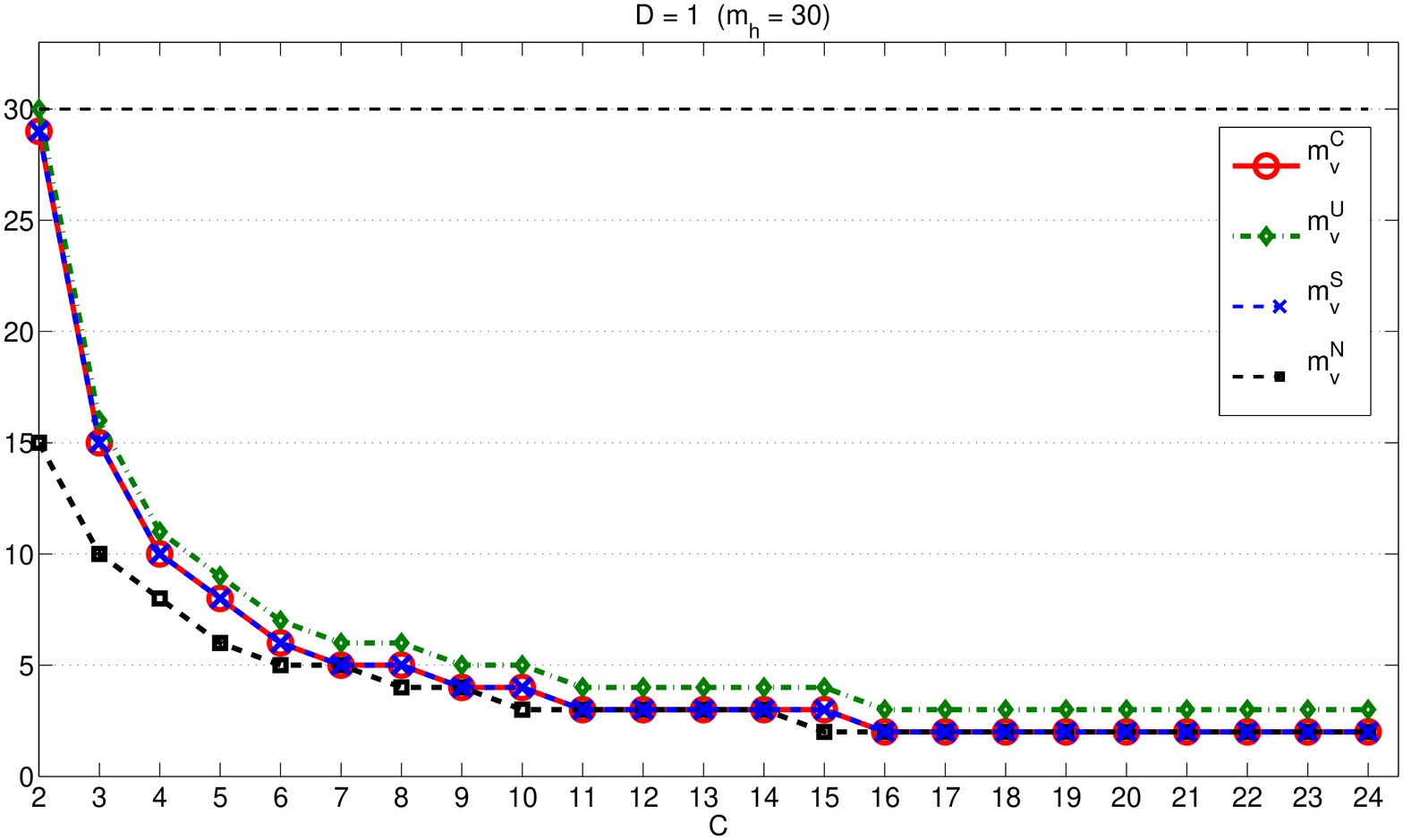}}
~
\subfloat[]{\includegraphics[trim=35mm 0mm 25mm 0mm, clip, width=0.5\textwidth]{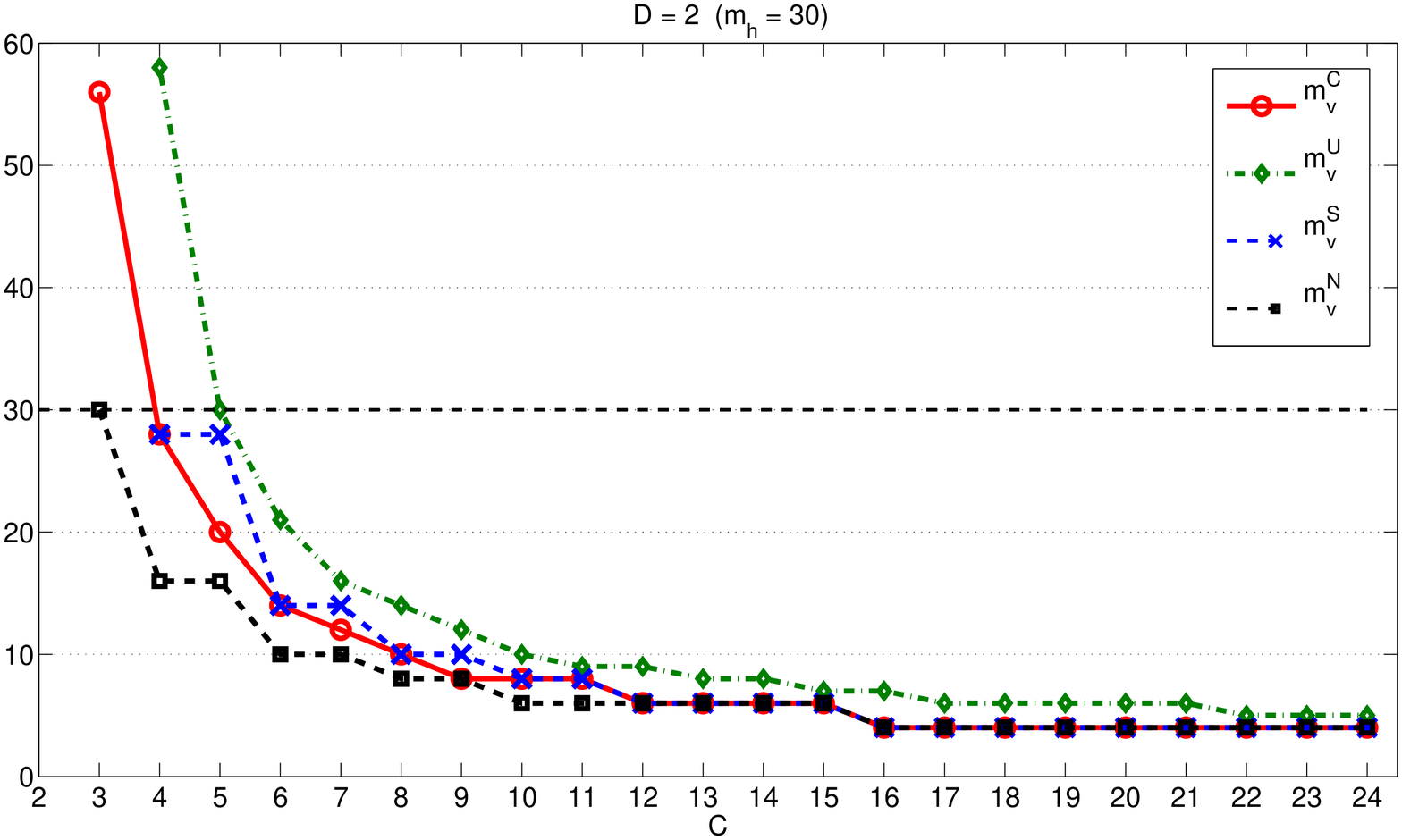}}
\end{minipage}
\begin{minipage}[b]{1\linewidth}
\centering
\vspace{-0.1in}
\subfloat[]{\includegraphics[trim=35mm 0mm 25mm 0mm, clip, width=0.5\textwidth]{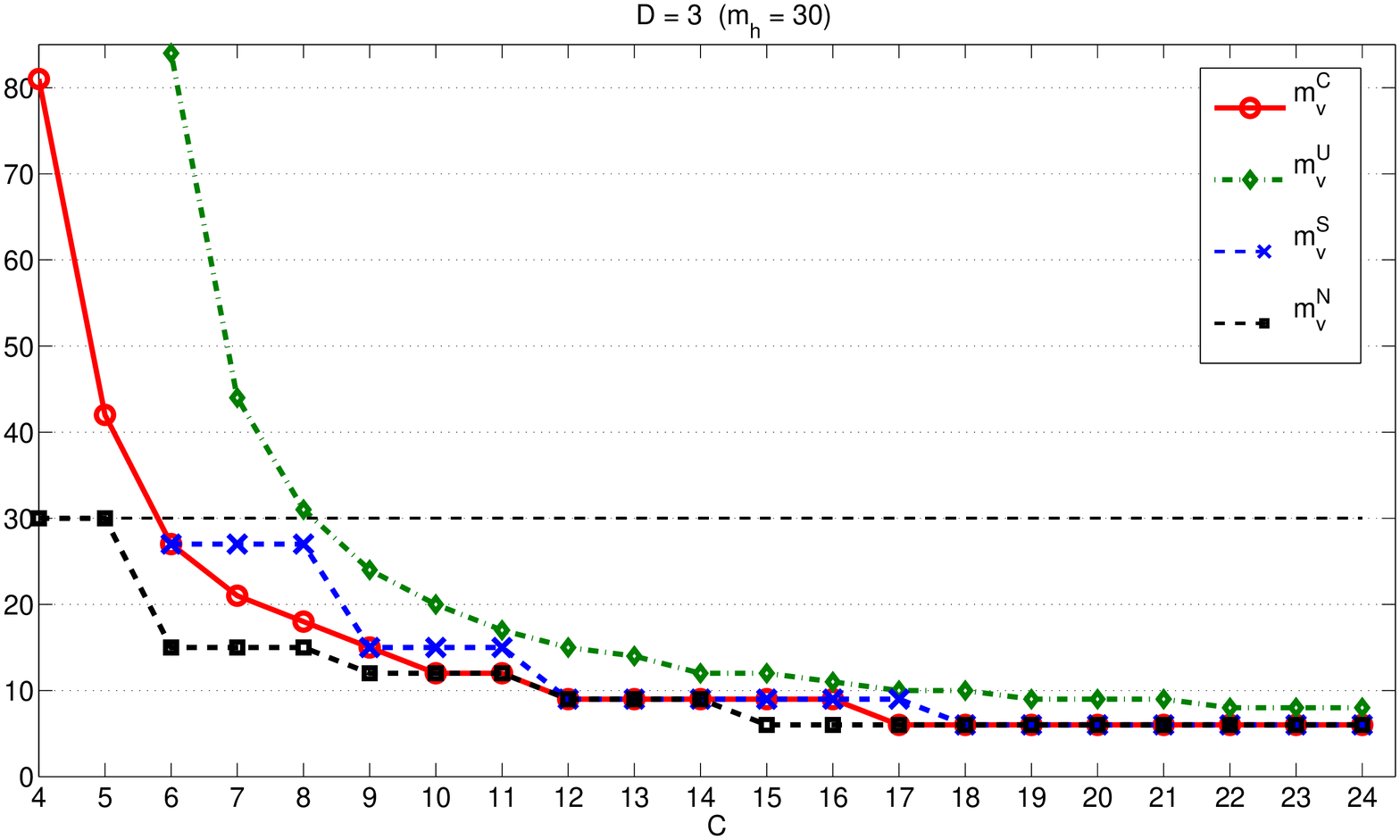}}
~
\subfloat[]{\includegraphics[trim=35mm 0mm 25mm 0mm, clip, width=0.5\textwidth]{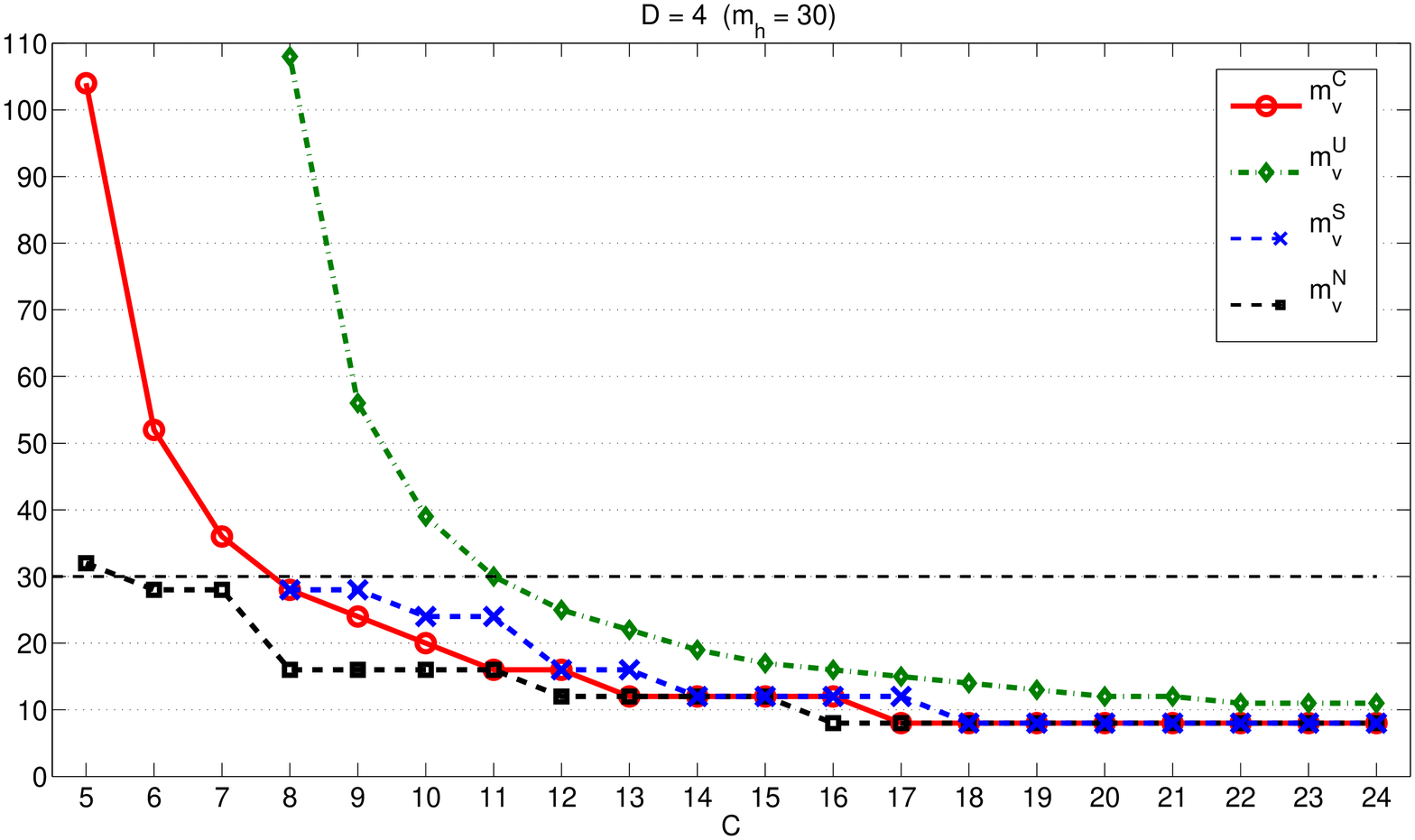}}
\end{minipage}
\vspace{-0.15in}
\caption{\small  The necessary ($m_v^{\st N}$), sufficient ($m_v^{\st S}$), and counting ($m_v^{\st C}$) filter lengths as function of the number of channels $C$, for an analysis filter length of $m_h = 30$ (marked by the dashed line in each panel) with subsampling factors of: (a) $D=1$; (b) $D=2$; (c) $D=3$; (d) $D=4$. The closed-form upper bound $m_v^{\st U}$ is also shown.}
\label{fig::mvbounds}
\vspace{-0.25in}
\end{figure}

\begin{figure}[t!]
\centering
\subfloat[]{\includegraphics[trim=20mm 0mm 25mm 0, clip, width=0.55\textwidth]{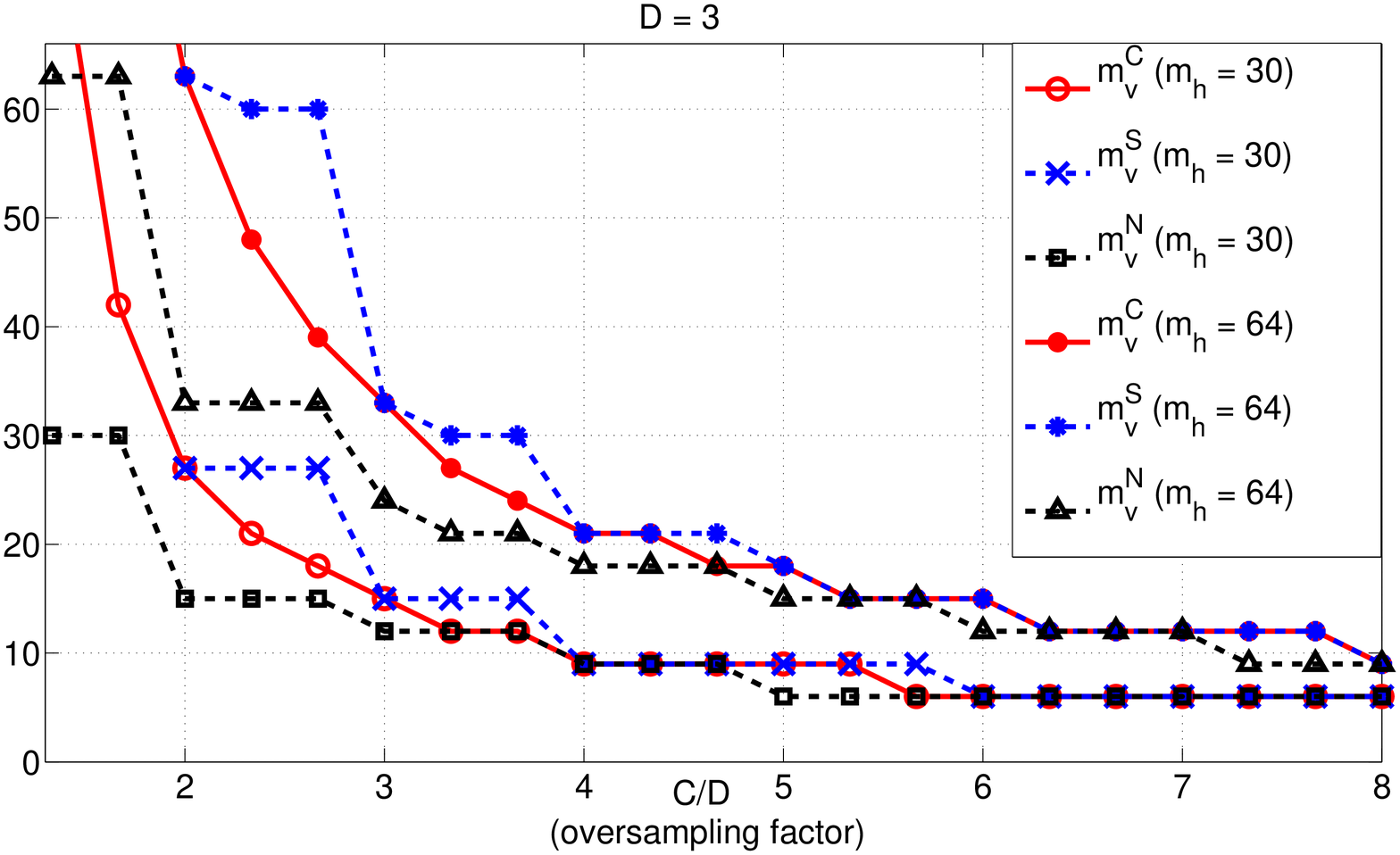}}
\subfloat[]{\includegraphics[trim=20mm 0mm 25mm 0, clip, width=0.47\textwidth]{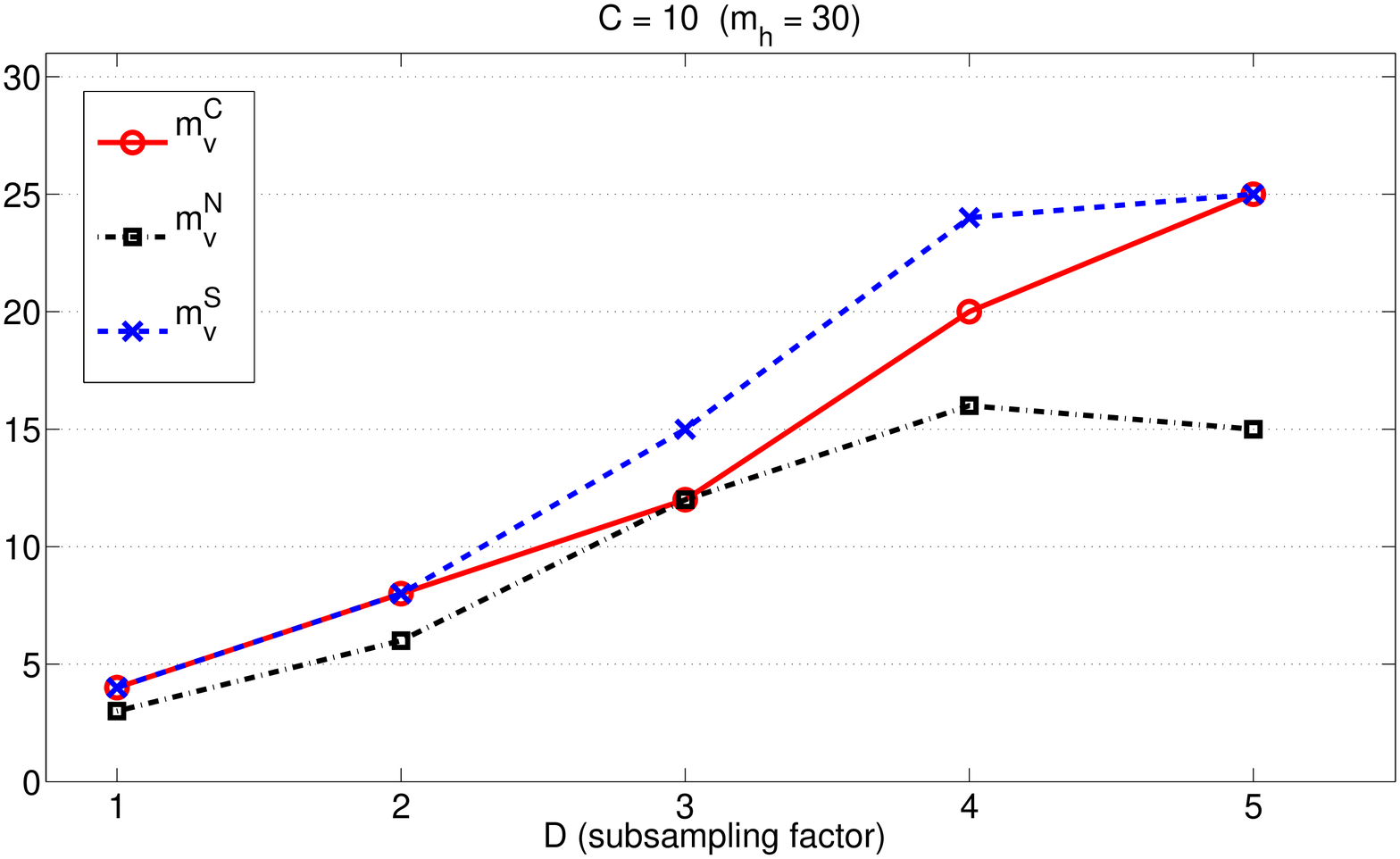}}
\vspace{-0.2in}
\caption{\small (a) Comparison of the necessary, sufficient, and counting $m_v$ filter lengths for two different analysis filters with lengths $m_h= 30$ and $m_h=64$, demonstrating that the gap between the filter lengths becomes negligible for moderately high oversampling factors; (b) Various $m_v$ filter lengths as a function of subsampling factor $D$ for a fixed number of channels $C=12$ and analysis filter length of $m_h = 30$.}
\label{fig::mvbounds2}
\vspace{-0.15in}
\end{figure}

Figure \ref{fig::mvbounds} shows the necessary, sufficient, and counting filter lengths as functions of the number of channels $C$, for an analysis filter length of $m_h = 30$ and subsampling factors of: (a) $D=1$; (b) $D=2$; (c) $D=3$; (d) $D=4$. The upper bound $m_v^{\st U}$ is also overlaid on the plots and the dashed line in each panel marks the analysis filter length $m_h$. Therefore, it is easy to verify the upper bound $\min(m_h, m_v^{\st U})$ in each panel. 
Next, Panel (a) of Fig.~\ref{fig::mvbounds2} plots the necessary, sufficient, and counting lengths for two different analysis filters lengths: $m_h= 30$ and $m_h=64$. The results demonstrate that the gap between the filter lengths becomes negligible for moderately high oversampling factors. In Panel (b), plots of the various $m_v$ filter lengths is provided as a function of subsampling factor $D$ for a fixed number of channels $C=12$ and analysis filter length of $m_h = 30$. The results suggest the the minimally required synthesis filter length for PR increases approximately proportional to the subsampling factor $D$, with a fixed number of channels and analysis filter length.

Together Figs. \ref{fig::mvbounds} and \ref{fig::mvbounds2} demonstrate that the gaps between the necessary, the sufficient, and the counting lengths are small. Further, these numerical results verify: (i) the upper bound $\min(m_h, m_v^{\st U})$ given in Proposition \ref{lem::corollary_one}; (ii) the relationships between the different filter lengths, also given in Proposition \ref{lem::corollary_one}.

\begin{figure}[t!]
\centering
\includegraphics[width=0.77 \textwidth]{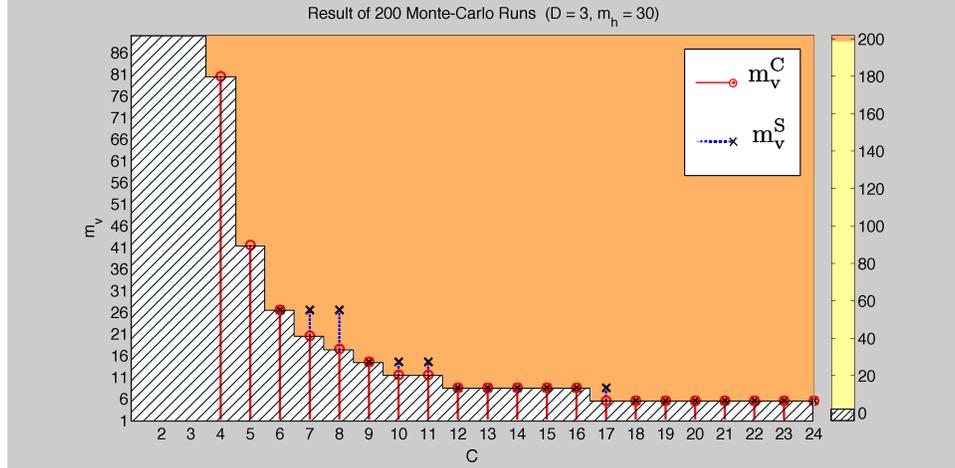}
\caption{\small Monte-Carlo simulation results, with randomly generated length $m_h=30$ analysis filters, for verification of the propositions with $C$ channels (horizontal axis) and $D$=$3$ subsampling. The color-bar indicates the number (out of 200) of Monte-Carlo runs  for which a PR synthesis bank of length $m_v$ (vertical axis) exists. The counting and sufficient lengths, $m_v^{\st C}$ and $m_v^{\st S}$, are overlaid on the graph as a function of $C$.}
\label{fig::prop_mcmc}
\vspace{-0.22in}
\end{figure}

Next, we move on to verification of the propositions. Figure \ref{fig::prop_mcmc} shows Monte-Carlo (M-C) simulation results for studying PR feasibility among randomly generated analysis FBs with 3-fold subsampling ($D$=$3$) and $C=1,\dots,24$ channels (horizontal axis). 
Each M-C run (from a total of 200) corresponded to generating an analysis FB comprising $C$ real-valued analysis filters of length $m_h = 30$, based on a uniform distribution with zero mean and variance of 100. 
The color-bar in the figure encodes the number of M-C runs (out of 200) for which a PR synthesis bank (allowing for multiple-of-$3$ delays) consisting of $C$ filters with length $m_v$ (vertical axis) exists. 
For each generated analysis FB, the computational process for determining numerically whether or not a PR synthesis FB exists involves checking the range condition in Lemma  \ref{thm::PRCond}  for all feasible choices of $n_0$. 
The counting and sufficient lengths, $m_v^{\st C}$ and $m_v^{\st S}$, are overlaid on the graph as a function of $C$.
As is seen from the figure, for $C\geq 2D=6$, all synthesis filter length choices satisfying $m_v\geq m_v^{\st S}$ resulted in 100\% success in achieving PR. This verifies  the claim of Proposition \ref{thm::FullRankGenericSuff}. The figure also shows that \emph{none} of synthesis banks with filter lengths below the counting length $m_v^{\st C}$ achieved PR (or delayed PR).
%
Moreover, since the necessary length $m_v^{\st N}$ bounds the counting length from below (Proposition \ref{lem::corollary_one}),  
the M-C results verify that none of the synthesis lengths $m_v$ that satisfied $m_v<m_v^{\st N}\leq m_v^{\st S}$ resulted in PR --- therefore, the minimal filter length that would allow for PR is at least as large as the derived necessary length. Consequently, the M-C results indirectly verify the claim of Proposition \ref{thm::FullRankGenericNec} as well.
%

The simulation results in Fig.~\ref{fig::prop_mcmc} suggest that the counting length $m_v^{\st C}(C,D,m_h)$ is both sufficient and necessary for feasibility of PR with a length-$m_v$ synthesis FB, i.e., the true minimum length $m_v^{\ast}$, introduced in Section \ref{sec::bounds_n_relations}, coincides with $m_v^{\st C}$. This provides grounds for the following conjecture.
(Note that, based on the bounds in Proposition \ref{lem::corollary_one}, this conjecture is consistent with the requirement that the true minimum length should lie in between the necessary and sufficient lengths.)
\vspace{-5pt}
\newtheorem*{conjecture_one}{Conjecture}
\begin{conjecture_one}\label{lem::conjecture_one}  
For a $D$-fold subsampled FIR analysis bank $\{ h_i\}_{i=1}^{\scsc C}$ with $m_h>D$, the following property holds generically\textsc{:} a  length-$m_v$ synthesis bank that achieves PR exists if and only if \mbox{$m_v\geq  m^{\st C}_v(C,D,m_h)$.} 
\end{conjecture_one}
\vspace{-3pt}

Regardless, as was pointed out in Section \ref{sec::bounds_n_relations} and verified in the results above, the gap between the counting length and our proven necessary and sufficient lengths is small, especially for moderately high oversampling factors.

\begin{figure}[t!]
\centering
\subfloat[]{\includegraphics[trim=25mm 0 25mm 0, clip, width=0.52\textwidth]{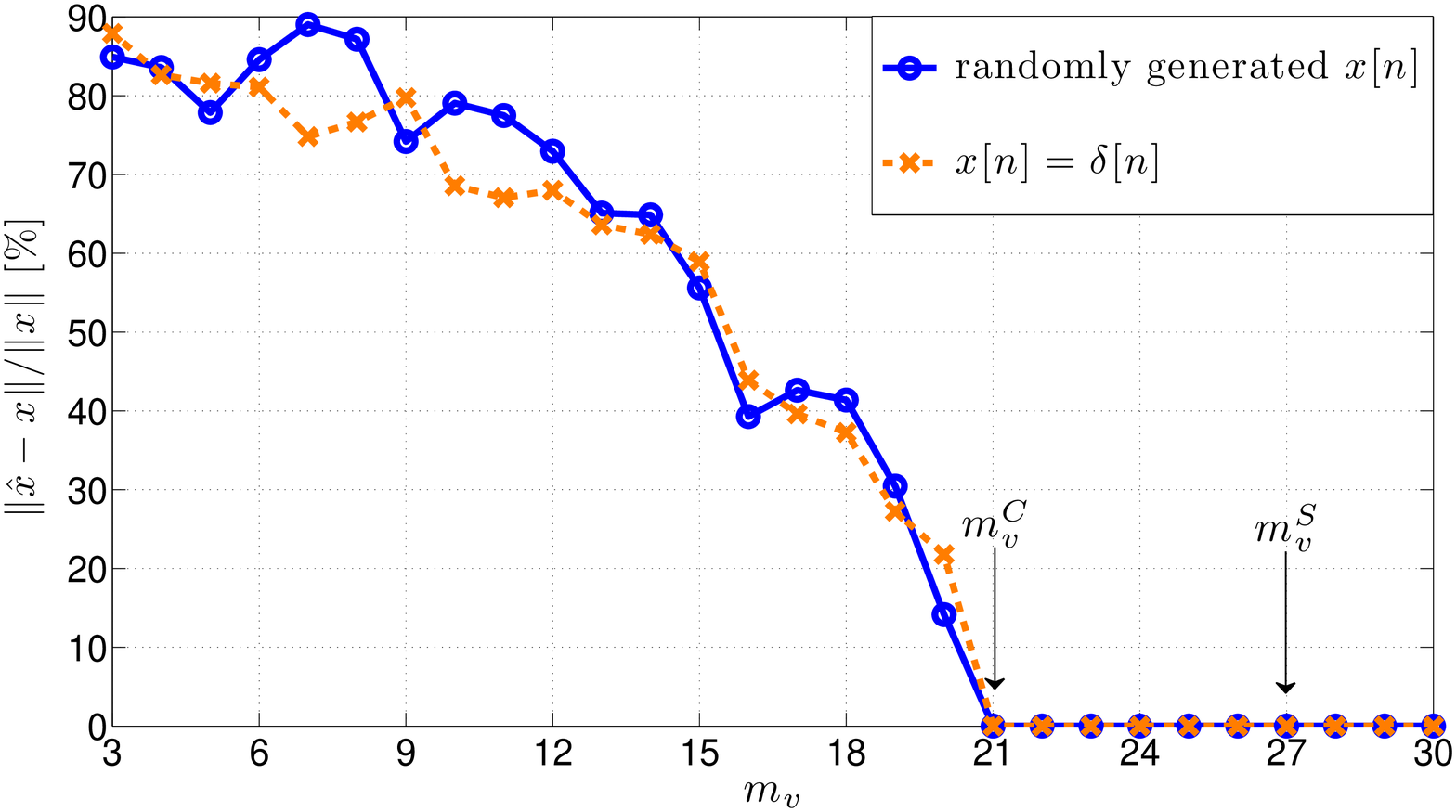}}
\subfloat[]{\includegraphics[trim=20mm 0 30mm 0, clip, width=0.5\textwidth]{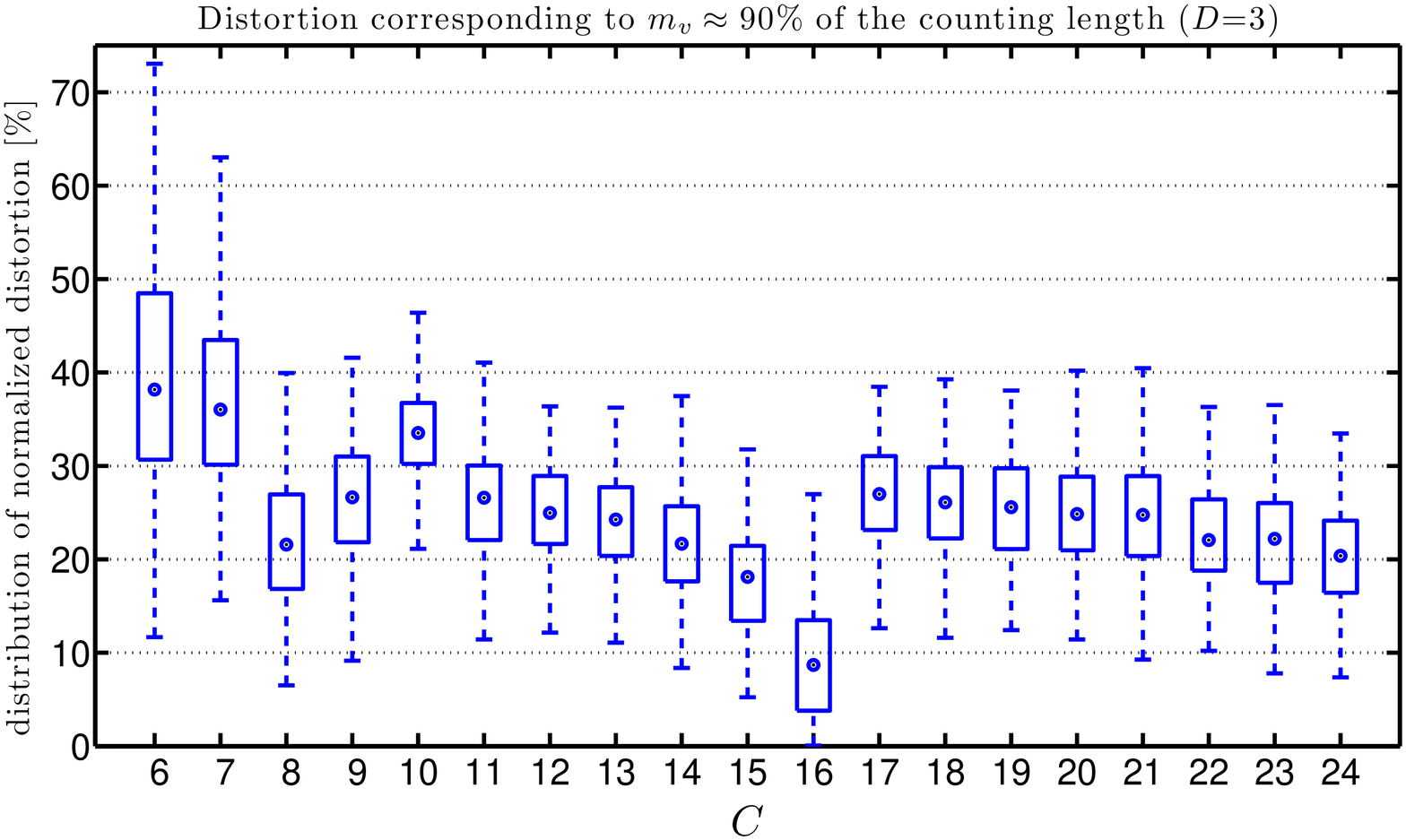}}
\vspace{-0.15in}
%
\caption{\small Signal distortion caused by synthesis filters that are too short.~(a) Normalized reconstruction distortion in percentage (noiseless scenario) as a function of $m_v$, the synthesis filter length, for a fixed (randomly generated) analysis FB with $D=3$ subsampling and filter length of $m_h=30$; results corresponding to two input signals, $x[n]$, are shown: one randomly generated with length $m_x=100$ (solid line) and one taken to be unit pulse (dashed line). 
(b) Boxplot showing the distribution of normalized distortion (in percentage) as a function of the number of channels $C$ corresponding to 200 Monte-Carlo runs with the following scenario: the $C$-channel analysis FB (with $D=3$) is generated randomly with a zero-mean i.i.d. unit-variance uniform distribution and reconstruction error corresponding to a synthesis FB of length $m_v=\lceil 0.9\times m_v^{\st C}(C,D,m_h)\rceil$ is computed.}
\label{fig::Distortion_BelowMinMv}
\vspace{-0.25in}
\end{figure}

In the last part of this section, we consider the feasibility of imperfect (near-perfect) reconstruction using synthesis filter lengths below those prescribed for PR by our theoretical results. To do so, assuming noise-free data, we need to compute the \emph{reconstruction error (distortion)} resulting from FIR synthesis with different synthesis filter lengths. That is, for each choice of $m_v$, we solve each of the $D$ (sampling-domain) polyphase matrix equations in \eqref{eq::polyphcond_time_genDelay} by applying $\bH_p^{\,\dagger}$ to both sides. Next, we collect all of the solutions to construct the synthesis bank, which enables computation of the reconstructed signal $\hat x[n]$ given the input $x[n]$. The resulting normalized reconstruction distortion (in percentage) is defined as $\Big\|\hat{x}[n] - x[n]\Big\|_2/\big\|x[n]\big\|_2 \times 100$.
Fig.~\ref{fig::Distortion_BelowMinMv}(a) shows this 
normalized reconstruction distortion as a function of $m_v$, the synthesis filter length, for a FB with $D=3$ subsampling and $m_h=30$ analysis filter length. Results corresponding to two input signals, $x[n]$, are shown: one randomly generated with length $m_x=100$ (solid line) and one taken to be the unit pulse (dashed line). As is seen from the figure, the reconstruction distortion is zero for $m_v$ values satisfying $m_v\geq m_v^{\st C} = 21$, which is consistent with the results in Fig.~\ref{fig::prop_mcmc}, but the distortion jumps to more than 10\% (for both inputs) when $m_v$ drops below $m_v^{\st C}$ and is more than 30\% with $m_v= m_v^{\st C} - 3 = 18$.

Finally, to study the effect of different analysis FB realizations on the resulting distortion, we conducted the following M-C simulation: normalized reconstruction distortions using a $C$-channel $3$-fold subsampled FIR synthesis FB were computed as a function of $C$ corresponding to $200$ Monte-Carlo runs, where in each run the analysis FB is generated randomly from a zero-mean i.i.d. unit-variance uniform distribution and the reconstruction distortion corresponding to a synthesis bank of length $m_v=\lceil 0.9\times m_v^{\st C}(C,D,m_h)\rceil$ is computed. The results are demonstrated in Panel (b) of Fig.~\ref{fig::Distortion_BelowMinMv} using the  \emph{boxplot} scheme \cite{McGTukLar78}. On each box, corresponding to a certain value for $C$, the central dot is the median of the distortion error percentage, and the edges of the box are the $25^\textrm{th}$ and $75\textrm{th}$ error percentiles. Each box has two \emph{whiskers}, which are a way to represent the range of variations of the error ``population'' whereby the upper and lower limits are extended to the most extreme data points not considered ``outliers'' \cite{McGTukLar78}.
Outliers\footnote{
Specifically, for each box, error values points are considered outliers if they are larger than $\q_3 + 1.5(\q_3 - \q_1)$ or smaller than $\q_1 - 1.5(\q_3 - \q_1)$, where $\q_1$ and $\q_3$ are the $25^{\mathrm{th}}$ and $75^{\mathrm{th}}$ percentiles of the error samples for the box, respectively. This range $\q_3 - \q_1$ is typically referred to as the inter-quartile range (also called the mid-spread) and the factor $1.5$ here is referred to as the whisker length.}
 are not shown. The maximum number of outliers was $5$ (corresponding to the box for $C=7$), i.e., the whiskers cover at least $97.5\%$ of the error values for each box.
As is seen from the figure, even a small deviation (here about $10\%$) from the minimally required filter lengths results in significant reconstruction distortion. (Note that for some FB realizations, the distortion can be quite small. However, this rarely happens: for the case of $C=16$, only $6\%$ of the generated FBs result in distortion of less than $1\%$.)

Overall, the results in Fig.~\ref{fig::Distortion_BelowMinMv} demonstrate that there would be little benefit in attempting to achieve non-perfect reconstruction to allow for synthesis lengths below those suggested by our propositions.

\section{Summary and Conclusions}\label{sec::disc_and_conc}

In this paper, we addressed the feasibility of PR using short FIR synthesis filters given an oversampled but otherwise general FIR analysis filter bank. We provided prescriptions for the shortest filter length of the synthesis bank that would guarantee PR. Our results are in the form of necessary and sufficient statements that hold generically, that is, only fail for contrived examples and pathological cases.  
For oversampling factors of at least two, we showed that our prescribed length for the synthesis filters is shorter than the analysis filters, decreases with increasing oversampling,  and is close to the derived necessary length for moderately high oversampling factors. Finally, using numerical studies, we demonstrated that choosing filter lengths that are only slightly below the prescribed regime results in significant signal distortion. 
%
%
%
The results have potential applications in synthesis FB design problems where the analysis bank is given, e.g., in multi-channel sensing/imaging systems where the channel characteristics are determined by the physics of the problem and cannot be fully manipulated. In addition, the presented work helps analyze and understand fundamental limitations in blind reconstruction of signals from data collected by unknown subsampled multi-channel systems.
A future area of research is to extend the results to higher dimensional signals and filter banks. 


\appendices 

\section{Proofs of Lemmas} 
\label{sec::app_proof_mvstar}

\subsection*{Proof of Lemma \ref{lem::FullRankGenericSuffRequired} } 

The sufficient-length condition in \eqref{eq::suffcond_condition} can be rewritten as 
\beqn
\sum_{p=0}^{D-1} \Bigg\lceil\frac{ \big\lceil (m_h - p)/D\big\rceil - 1}{\lfloor m_v/D \rfloor}\Bigg\rceil  \leq C-D < D,
\label{eq::suffcond_condition_lemmaproof}
\eeqn
where the right-most inequality holds for $C<2D$. However, each of the $D$ terms in the  sum on left-hand side is at least 1 since $m_h\geq 2D$. Therefore, the sum is greater than or equal to $D$, which is a contradiction. \hfill $\square$

\subsection*{Proof of Lemma \ref{lem::ConstructionIdea}  } 


The proof is by contradiction. Assume $A$ is not full row rank. Therefore, there should exist a row in  $A$, say, the $m$-th row, that is a linear combination of the other rows. By Property (ii) in the statement of the lemma, this row has at least one nonzero element, say, the $n$-th element. Consider this ($m,n$)-th entry:  $A[m,n]\neq 0$. According to Property (i) stated in the lemma, no other nonzero element exists in the $n$-th column of the matrix. Therefore, $A$ is a linear combination of zero elements and is therefore zero. This is a contradiction with $A[m,n]\neq 0$; hence, the result. 
\hfill $\square$

\subsection*{Proof of Lemma \ref{lem::ConstructionRowCovering}   }


First, note that the problem can considered independently for each of the block-rows since each Toeplitz block in $\bH_p$ only belongs to one block-row. 
It is easy to see that the lemma's claim is equivalent to asserting that, using Algorithm \ref{alg::one}, among the 1-diagonal assignments that cover the rows in the $\ell$-th block-row \emph{only the last one} can be of the nontrivial type shown in Fig.~\ref{fig::AlgCases_SuffProof}(b). To prove this, note that assigning $1$ to the last free entry (i.e., the free entry with the highest row index) in the first column of a Toeplitz block means that the bottom-right entry of the block (corresponding to the last row and last column in the block, which is never a structural zero) would be assigned $1$ as well. Therefore, the last row of the block-row is assigned a nonzero entry, i.e., the entire block-row is covered. 
\hfill $\square$
%

\subsection*{Proof of Lemma \ref{lem::Line2_Alg2}  } 

The necessary length defined in \eqref{eq::neccond_condition} is an integer functional from $\mathbb{N}^3$ onto $\mathbb{N}$:
\beqn
m^{\st N}_v(C,D,m_h) = \min \bigg\{ m_v\in\mathbb{N} \:\bigg\lvert\: \frac{C}{D} \geq 1 + \frac{1}{D}\sum_{p=0}^{D-1} \bigg\lfloor\frac{m_{h;p}  - 1}{\lfloor m_v/D \rfloor}\bigg\rfloor  \:\bigg\}
\label{eq::neccond_condition_appA}
\eeqn
A property of this functional is that $m^{\st N}_v(C,D,m_h)\leq m_h$ (given  $C/D \geq 2$ and $m_h> D$), which is proven as part of the proof of Proposition \ref{lem::corollary_one} in this appendix.  
Therefore, the assumption $m_v<m^{\st N}_v(C,D,m_h)$ implies $m_v<m_h$. We have
\beqn
m_{h;0}=\left\lceil\frac{m_h}{D}\right\rceil\geq\frac{m_h}{D}>\frac{m_v}{D}\geq\left\lfloor\frac{m_v}{D}\right\rfloor = m_{v;\po},
\nonumber
\eeqn
where, in the last step, we used Property (c) in \eqref{eq::polyphlength_properties}.
\hfill $\square$

\subsection*{Proof of Lemma \ref{thm::CountingCond}  } 

For $p$-th sampling-domain polyphase condition in \eqref{eq::polyphcond_time_genDelay}, a necessary condition for the matrix $\bH_p$ to be full row rank is for its number of rows to be less than or equal to its number of columns:
\beqnn
m_{h} + Dm_{v;p} - D \leq C \: m_{v;p}  \;\Rightarrow\; m_{v;p} \geq \frac{m_h - D}{C-D} \;\Rightarrow\; m_{v;p} \geq  \left\lceil  \frac{\obliquefrac{m_h}{D} - 1}{\obliquefrac{C}{D} - 1} \right\rceil .
\eeqnn
This inequality should hold for all $m_{v;p}$ where $p=0,\dots,D-1.$ Therefore, it should hold for\\ $\min_{(p=0,\dots,D-1)} m_{v;p} = \lfloor m_v/D \rfloor,$ where we used the definition of $m_{v;p}$ described in Section \ref{sec::preliminary_polyph}. Therefore, we have
\beqn
\left\lfloor \frac{m_v}{D} \right\rfloor  \geq \left\lceil \frac{\obliquefrac{m_h}{D} - 1}{\obliquefrac{C}{D} - 1} \right\rceil.
\label{eq::LemmaCountingCondProof}
\eeqn
Now, it is easy to check that $m^{\st C}_v(C,D,m_h) \defeq D\Big\lceil (\frac{m_h}{D} - 1)/(\frac{C}{D} - 1)\Big\rceil$ satisfies \eqref{eq::LemmaCountingCondProof}, whereas \mbox{$m^{\st C}_v(C,D,m_h) -1$} does not. Therefore, $m^{\st C}_v(C,D,m_h)$ is the smallest integer that satisfies the condition in \eqref{eq::LemmaCountingCondProof}. 
\hfill $\square$




\smallskip
\subsection*{Proof of Proposition \ref{lem::corollary_one} }

The results in Proposition \ref{lem::corollary_one} follow from combining those in the following two lemmas. 
Lemma \ref{lem::FullRankGenericSuffFolk} provides a set of properties including a closed-form upper bound for the sufficient length $m_v^{\st S}$:  
\vspace{-0.1in}
\newtheorem{lemma_suff_folk}[lemma]{Lemma}
\begin{lemma_suff_folk}\label{lem::FullRankGenericSuffFolk}  
If $C/D \geq 2$ and $m_h> D$, then $m^{\st S}_v(C,D,m_h)$  has the following properties:
\begin{itemize}
\item[(i)] $m_v^{\st S} = m_v^{\st C}$ if $C=kD$ for integers $k\geq 2$.
\item[(ii)] 
$m_v^{\st C} \leq m_v^{\st S} \leq \min(m_h, m_v^{\st U})$, where $\displaystyle m_v^{\st U}(C,D,m_h) = D + \bigg\lceil\frac{m_h-D}{(C+1)/D-2}\bigg\rceil$.
\end{itemize}
\end{lemma_suff_folk}

\noindent Lemma \ref{lem::FullRankGenericNecFolk} describes the relationship between the necessary, sufficient, and counting filter lengths:
%
%
\vspace{-0.05in}
\newtheorem{lemma_nec_folk}[lemma]{Lemma}
\begin{lemma_nec_folk}\label{lem::FullRankGenericNecFolk}  
If $C/D \geq 2$ and $m_h> D$, then $m^{\st N}_v(C,D,m_h)$  has the following property:
\beqn
\max\big(D,m_v^{\st L}\big)\leq m_v^{\st N} \leq m_v^{\st C}\leq m_v^{\st S}, \textrm{ where } m_v^{\st L}(C,D,m_h) = \Bigg\lceil\frac{m_h-D}{C/D}\Bigg\rceil .
\nonumber
\eeqn
\end{lemma_nec_folk}

\noindent  Proofs for Lemmas \ref{lem::FullRankGenericSuffFolk}  and  \ref{lem::FullRankGenericNecFolk} are included below. 
\medskip

\begin{IEEEproof}[Proof of Lemma  \ref{lem::FullRankGenericSuffFolk}]
%
\textsc{Part} (i).
First, let us show that for a real positive number $\alpha$ and an integer $M$, $\lceil \alpha/M \rceil \geq \lceil \alpha \rceil / M.$ This trivially holds for $\alpha \in \mbb{N}$. To show this for $\alpha \notin \mbb{N}$, define $[\alpha]_1 = \alpha - \lfloor \alpha \rfloor,$ and write $\lfloor \alpha \rfloor = m_{\alpha} M + r_{\alpha}$ where the remainder satisfies  $0\leq r_{\alpha} \leq M-1$. Therefore, the left-hand side of the inequality is $\lceil \alpha/M \rceil = m_{\alpha} + \Big\lceil\big(r_{\alpha} + [\alpha]_1\big)/M\Big\rceil = m_{\alpha} + 1,$ where we used $r_{\alpha} + [\alpha]_1 \leq M-1 + [\alpha]_1 < M$. Similarly, the right-hand side can be written as: $\lceil \alpha \rceil/M = \big(\lfloor \alpha \rfloor + 1\big)/M = m_{\alpha} + \big(r_{\alpha}+1\big)/M \leq  m_{\alpha} + 1$, which completes the proof.  
Now, consider a synthesis filter length $m_v^{\ast}$ that satisfies the ``counting'' condition given in \eqref{eq::lemma_counting} (in Lemma \ref{thm::CountingCond}). By definition, $m_v^{\ast} \geq m_v^{\st C}$. With $C/D = k \in\mbb{N}$, we have that:
\beqn 
\left\lfloor \frac{ m_v^{\ast} }{D} \right\rfloor \geq \left\lceil \frac{m_h/D - 1}{k-1} \right\rceil \geq \frac{\big\lceil m_h/D -1\big\rceil}{k-1} \:\Rightarrow\:  \frac{\big\lceil m_h/D -1\big\rceil}{\big\lfloor  m_v^{\ast} /D \big\rfloor} \leq k - 1,
\nonumber
\eeqn
where we used the property proved above with $\alpha = m_h/D - 1$ and $M=k-1$; also $\lfloor m_v^{\ast} /D\rfloor >0$ since it is assumed that $m_h>D$. Consequently, we have:
\beqn 
 \frac{\big\lceil m_h/D -1\big\rceil}{\big\lfloor  m_v^{\ast}/D \big\rfloor} \leq k - 1 \:\Rightarrow\:  
  \frac{\big\lceil m_h/D\big\rceil -1}{\big\lfloor  m_v^{\ast}/D \big\rfloor} \leq k - 1
  \Rightarrow\:  
    \frac{\Big\lceil \big(m_h - p\big)/D\Big\rceil -1}{\big\lfloor  m_v^{\ast}/D \big\rfloor} \leq k - 1
    \qquad (0\leq p \leq D-1).
\nonumber
\eeqn
Hence,
\beqn
\sum_{p=0}^{D-1} \Bigg\lceil\frac{ \big\lceil (m_h - p)/D\big\rceil - 1}{\lfloor  m_v^{\ast}/D \rfloor}\Bigg\rceil  \leq (k-1)D = C-D,
\label{eq::suffcond_countingcondequalcase}
\eeqn
since $C=kD.$
Comparing \eqref{eq::suffcond_condition_lemmaproof} and \eqref{eq::suffcond_countingcondequalcase}, it is clear that any $m_v^{\ast}$ satisfying the counting condition in \eqref{eq::lemma_counting} would also satisfy the generic sufficient condition in Proposition \ref{thm::FullRankGenericSuff}. 
Hence, based on the definition of $m_v^{\st S}$, it follows that\footnote{This is equivalent to saying that: $\min\limits_{m_v\in S_1} m_v \leq \min\limits_{m_v\in S_2} m_v$, when $S_1\supseteq S_2$.
}
 $m_v^{\st S}\leq m_v^{\st C}$ (for the case of $C=kD$). On the other hand, as will be shown below,  $m_v^{\st S} \geq m_v^{\st C}$ (in general). This concludes the proof of $m_v^{\st S} = m_v^{\st C}$ (given $C=kD$).
\medskip

\noindent \textsc{Part} (ii). The proof has three parts as given below.\\
\noindent (ii--1) Proof of $m_v^{\st C} \leq m_v^{\st S}$. ~ 
Based on the proof of Proposition \ref{thm::FullRankGenericSuff}, it is clear that the condition for $m_v$ in the proposition is a sufficient condition for all $\bH_p$ ($p=0,\dots,D-1)$ to be \emph{generically} full rank (which in turn implies PR for the filter bank). In the proof, a particular matrix construction for $\bH_p$ is provided that is shown to have full row rank.
Now, a \emph{necessary} condition for all $\bH_p$ to have full row rank is given in Lemma \ref{thm::CountingCond}.
Consequently, since $m_v^{\st C}$ is, by definition, the minimum $m_v$ satisfying that condition, we have: $m_v^{\st S} \geq m_v^{\st C}$.

\noindent (ii--2)~Proof of $m_v^{\st S}\leq m_h$. ~ 
With $C\geq2D$, we have: $D\leq C-D$. 
Therefore, to prove $m_v^{\st S} \leq m_h$ all we need to show is that \eqref{eq::suffcond_condition_lemmaproof} holds for the choice of $m_v = m_h$. For $p=0,\dots,D-1$, we have
\beqn
\bigg\lceil \frac{m_h - p}{D} \bigg\rceil - 1  ~ \leq ~ \Big\lceil \frac{m_h}{D} \Big\rceil - 1  \leq  \Big\lfloor \frac{m_h}{D} \Big\rfloor 
\Rightarrow 
\frac{ \big\lceil (m_h - p)/D\big\rceil - 1}{\lfloor m_h/D \rfloor} \leq  1 
\Rightarrow 
\Bigg\lceil\frac{ \big\lceil (m_h - p)/D\big\rceil - 1}{\lfloor m_h/D \rfloor}\Bigg\rceil  \leq 1, \nonumber 
\eeqn
where the second step follows from the assumption of $m_h> D$ (as stated in the lemma). Summing this up for all $p=0,\dots,D-1,$ completes the proof:
\beqn
\sum_{p=0}^{D-1}  \Bigg\lceil\frac{ \big\lceil (m_h - p)/D\big\rceil - 1}{\lfloor m_h/D \rfloor}\Bigg\rceil  \leq  D ~ \leq ~ C - D. \nonumber
\eeqn

\noindent (ii--3) Proof of $m_v^{\st S}\leq m_v^{\st U}$. ~ 
Note that we can bound the left-hand side of lemma's condition \eqref{eq::suffcond_condition_lemmaproof} as follows:
\beqn
\sum_{p=0}^{D-1}  \Bigg\lceil\frac{ m_{h;p} - 1}{\lfloor m_v/D \rfloor}\Bigg\rceil  < D + \sum_{p=0}^{D-1}  \frac{ m_{h;p}  - 1}{\lfloor m_v/D \rfloor} &=& D + \frac{1}{\lfloor m_v/D \rfloor} \sum_{p=0}^{D-1} \Big(m_{h;p} -1\Big) \nonumber \\
&=& D + \frac{m_h-D}{\lfloor m_v/D \rfloor} < D + \frac{m_h - D}{m_v/D - 1},
\label{eq::suffcond_upperbound}
\eeqn
where we used the identity $ \sum_{p=0}^{D-1}   m_{h;p} = m_h$.
The following shows that if $m_v$ is chosen such this upper bound is no more than $C-D+1$, then the sufficient-length condition, restated in \eqref{eq::suffcond_condition_lemmaproof}, would be satisfied:
\beqn
\underbrace{D + \frac{m_h - D}{m_v/D - 1}  \leq C-D+1}_{\displaystyle \textrm{(}\ast\textrm{)}} & \Rightarrow & \sum_{p=0}^{D-1}  \Bigg\lceil\frac{ m_{h;p} - 1}{\lfloor m_v/D \rfloor}\Bigg\rceil < D + \frac{m_h - D}{m_v/D - 1} \leq C-D+1 \nonumber \\
& \Rightarrow & \sum_{p=0}^{D-1}  \Bigg\lceil\frac{ m_{h;p} - 1}{\lfloor m_v/D \rfloor}\Bigg\rceil \leq C-D. \nonumber
\eeqn
Note that if $m_v^{\st U}\in\mbb{N}$ satisfies ($\ast$), then based on the definition of $m_v^{\st S}$ in Section \ref{sec::SuffCond_statement} we would have $m_v^{\st S}\leq m_v^{\st U}$. All that remains to be shown is that $m_v^{\st U}$ given in the statement of the lemma satisfies ($\ast$):
\beqn
m_v^{\st U} = D + \left\lceil \frac{m_h-D}{(C+1)/D-2}\right\rceil  \Rightarrow  m_v^{\st U} \geq \frac{m_h+C-3D+1}{(C+1)/D-2} 
 \Rightarrow  m_h + C - 3D + 1 \leq \frac{m_v^{\st U}}{D}\Big(C + 1 - 2D \Big) \nonumber \\
 \Rightarrow   m_h - D \leq \Big(\frac{m_v^{\st U}}{D} - 1 \Big) \Big(C + 1 - 2D \Big) \Rightarrow  D + \frac{m_h - D}{m_v^{\st U}/D - 1}  \leq  C-D+1, \nonumber
\eeqn
where we used the assumptions $C\geq 2D$ and $m_h> D$. This completes proof of Lemma \ref{lem::FullRankGenericSuffFolk}. 
\end{IEEEproof}
\bigskip

\begin{IEEEproof}[Proof of Lemma  \ref{lem::FullRankGenericNecFolk}]
%
%
We first show the lower bound: $m_v^{\st N}\geq \max\big(D, m_v^{\st L}\big)$, where $m_v^{\st L} = \big\lceil\frac{m_h - D}{C/D}\big\rceil$. The inequality $m_v^{\st N}\geq D$ is trivial. 
Rewriting \eqref{eq::neccond_condition}, we have:
\beqn
C-D \geq \sum_{p=0}^{D-1}  \Bigg\lfloor\frac{ m_{h;p} - 1}{\lfloor m_v^{\st N}/D \rfloor}\Bigg\rfloor > \sum_{p=0}^{D-1}  \bigg(\frac{ m_{h;p} - 1}{\lfloor m_v^{\st N}/D \rfloor} - 1\bigg) = -D + \frac{1}{\lfloor m_v^{\st N}/D \rfloor}\Big(\sum_{p=0}^{D-1} m_{h;p} - 1\Big)  = -D + \frac{m_h - D}{\lfloor m_v^{\st N}/D \rfloor} \nonumber \\
\Rightarrow  C\geq \frac{m_h - D}{\lfloor m_v^{\st N}/D \rfloor}   \Rightarrow  \frac{\lfloor m_v^{\st N}/D \rfloor}{m_h - D} \geq \frac{1}{C}     \Rightarrow     \frac{m_v^{\st N}}{D} \geq \Big\lfloor \frac{m_v^{\st N}}{D} \Big\rfloor \geq \frac{m_h - D}{C} \Rightarrow  m_v^{\st N}\geq \frac{m_h - D}{C/D}  \Rightarrow   m_v^{\st N}\geq \Big\lceil\frac{m_h - D}{C/D} \Big\rceil. 
\nonumber
\eeqn

The inequality $m_v^{\st C}\leq m_v^{\st S}$ was already shown in Lemma \ref{lem::FullRankGenericSuffFolk}, Part (ii).
In the following we prove that $m_v^{\st C} \leq m_v^{\st N}$. 
For any integer $m_v\geq D$,  it follows that:
\beqn
\sum_{\ell=0}^{D-1} \Bigg\lfloor\frac{m_{h;\ell}  - 1}{\lfloor m_v/D \rfloor}\Bigg\rfloor & \leq &
\sum_{\ell=0}^{D-1} \frac{m_{h;\ell}  - 1}{\lfloor m_v/D \rfloor}
= \frac{-D + \sum_{\ell=0}^{D-1}  m_{h;\ell}}{\lfloor m_v/D \rfloor} 
=
\frac{m_h - D}{\lfloor m_v/D \rfloor}. 
\label{eq::neccond_appendix_eq1}
\eeqn
Next, assume an integer $m_v$ 
satisfies the counting condition given in \eqref{eq::lemma_counting}, i.e.,   $m_v\geq  m_v^{\st C}$. We have:
\beqn
\left\lfloor \frac{m_v}{D} \right\rfloor  \geq \left\lceil \frac{\obliquefrac{m_h}{D} - 1}{\obliquefrac{C}{D} - 1} \right\rceil \Rightarrow 
\left\lfloor \frac{m_v}{D} \right\rfloor  \geq \frac{\obliquefrac{m_h}{D} - 1}{\obliquefrac{C}{D} - 1} \Rightarrow 
C-D  \geq\frac{m_h - D}{ \left\lfloor m_v/D \right\rfloor },
\label{eq::neccond_appendix_eq2}
\eeqn
where we used $m_v\geq D$, which is implied from $m_h> D$ and \eqref{eq::lemma_counting}. 
Combining \eqref{eq::neccond_appendix_eq1} and \eqref{eq::neccond_appendix_eq2}, we have:
\beqn
\sum_{\ell=0}^{D-1} \Bigg\lfloor\frac{m_{h;\ell}  - 1}{\lfloor m_v/D \rfloor}\Bigg\rfloor \leq C-D.
\label{eq::neccond_appendix_eq3}
\eeqn
This means that $m_v$ satisfies the necessary length condition of Proposition  \ref{thm::FullRankGenericNec} given in \eqref{eq::neccond_condition}, which by the definition of $m_v^{\st N}$, implies that $m_v^{\st N}\leq m_v \leq m_v^{\st C}$. 
 This completes the proof of Lemma \ref{lem::FullRankGenericNecFolk}. 
%
\end{IEEEproof}

%

\subsection*{Proof of Corollary \ref{lem::GapBounds}} 


Applying Lemmas \ref{lem::FullRankGenericSuffFolk} and \ref{lem::FullRankGenericNecFolk}, it follows that:
\beqn
\Gamma_{\st S|\st N}(C,D,m_h) &\leq&  m_v^{\st U}(C,D,m_h) - m_v^{\st L}(C,D,m_h)  \nonumber \\ 
&\!\!\!\!\leq\!\!\!\!&  1 + D + \frac{m_h-D}{\frac{C+1}{D}-2} - \frac{m_h-D}{C/D} = 
1+D+ \frac{m_h-D}{\frac{C}{D}\Big( \frac{C}{2D-1} - 1\Big)} <  1+D + \frac{2 m_h}{\frac{C}{D}\Big( \frac{C}{D-0.5} - 2\Big)}. \nonumber
\eeqn
where we used the assumption $C/D\geq 2$.

Applying \eqref{eq::lemma_countingBounds}  and definitions of $m_v^{\st U}$ and $m_v^{\st L}$ given in Lemmas \ref{lem::FullRankGenericSuffFolk} and \ref{lem::FullRankGenericNecFolk}, respectively, we have:
\beqn
\Gamma_{\st U|\st C}(C,D,m_h) &<&  \Big(1 + D + \frac{m_h-D}{\frac{C+1}{D}-2}\Big) - \Big(\frac{m_h - D}{\frac{C}{D} - 1}\Big)  < 1 + D + \frac{m_h}{\big(\frac{C}{D} - 1\big) \big(\frac{C-1}{D-1} - 2\big)}. \nonumber\\
\Gamma_{\st C|\st L}(C,D,m_h) &<&  \Big(1 + D + \frac{m_h-D}{C/D-1}\Big) - \Big(\frac{m_h-D}{C/D}\Big) < 1 + D + \frac{m_h}{\frac{C}{D}\big( \frac{C}{D} - 1\big)},
\nonumber
\eeqn
where we applied the assumptions $D\geq 2$ and $C/D\geq 2$. 
\hfill $\square$



\bibliographystyle{IEEEtran}
\bibliography{IEEEabrv,rm}

\end{document}